\documentclass[numberedappendix,iop]{emulateapj}
\usepackage{apjfonts}
\usepackage{array}
\usepackage[usenames]{color}
\begin{document}
%\submitted{Final version \today}
\newcommand{\comment}[1]{}
\newcommand{\risa}[1]{\textcolor{red}{(\bf #1)}}
\newcommand{\michael}[1]{\textcolor{blue}{(\bf #1)}}
\definecolor{purple}{RGB}{160,32,240}
\newcommand{\peter}[1]{\textcolor{purple}{(\bf #1)}}
\newcommand{\macc}{M_\mathrm{acc}}
\newcommand{\mpeak}{M_\mathrm{peak}}
\newcommand{\mnow}{M_\mathrm{now}}
\newcommand{\vacc}{v_\mathrm{acc}} %_\mathrm{max}}
\newcommand{\vpeak}{v_\mathrm{peak}} %_\mathrm{max}}
\newcommand{\vnow}{v^\mathrm{now}_\mathrm{max}}

\newcommand{\Mnfw}{M_\mathrm{NFW}}
\newcommand{\Msun}{M_{\odot}}
\newcommand{\mvir}{M_\mathrm{vir}}
\newcommand{\rvir}{R_\mathrm{vir}}
\newcommand{\vmax}{v_\mathrm{max}}
\newcommand{\vmac}{v_\mathrm{max}^\mathrm{acc}}
\newcommand{\mvac}{M_\mathrm{vir}^\mathrm{acc}}
\newcommand{\sfr}{\mathrm{SFR}}
\newcommand{\plotgrace}[1]{\includegraphics[angle=-90,width=\columnwidth,type=eps,ext=.eps,read=.eps]{#1}}
\newcommand{\plotgraceflip}[1]{\includegraphics[angle=-90,width=\columnwidth,type=eps,ext=.eps,read=.eps]{#1}}
\newcommand{\plotlargegrace}[1]{\includegraphics[angle=-90,width=2\columnwidth,type=eps,ext=.eps,read=.eps]{#1}}
\newcommand{\plotlargegraceflip}[1]{\includegraphics[angle=-90,width=2\columnwidth,type=eps,ext=.eps,read=.eps]{#1}}
\newcommand{\plotminigrace}[1]{\includegraphics[angle=-90,width=0.5\columnwidth,type=eps,ext=.eps,read=.eps]{#1}}
\newcommand{\plotmicrograce}[1]{\includegraphics[angle=-90,width=0.25\columnwidth,type=eps,ext=.eps,read=.eps]{#1}}
\newcommand{\plotsmallgrace}[1]{\includegraphics[angle=-90,width=0.66\columnwidth,type=eps,ext=.eps,read=.eps]{#1}}

\newcommand{\hinv}{h^{-1}}
\newcommand{\mpc}{\rm{Mpc}}
\newcommand{\hmpc}{$\hinv\mpc$}

\shortauthors{BEHROOZI ET AL}
\shorttitle{Mergers and Mass Accretion End Well Outside Clusters}

\title{Mergers and Mass Accretion for Infalling Halos Both End Well Outside Cluster Virial Radii}
\author{Peter S. Behroozi\altaffilmark{1,2}, Risa H. Wechsler\altaffilmark{2}, Yu Lu\altaffilmark{2}, Oliver Hahn\altaffilmark{3}, Michael T. Busha\altaffilmark{2,4}, Anatoly Klypin\altaffilmark{5}, Joel R. Primack\altaffilmark{6}}
\altaffiltext{1}{Space Telescope Science Institute, Baltimore, MD 21218 USA} 
\altaffiltext{2}{Physics Department, Stanford University; Department of Particle and Particle Astrophyiscs, 
SLAC National  Accelerator Laboratory; Kavli Institute for Particle Astrophysics and Cosmology
Stanford, CA 94305 USA} 
\altaffiltext{3}{Institute for Astronomy, ETH Zurich, CH-8093 Zurich, Switzerland}
\altaffiltext{4}{Institute for Theoretical Physics, University of Zurich, 8006 Zurich, Switzerland}
\altaffiltext{5}{Astronomy Department, New Mexico State University, Las Cruces, NM, 88003 USA}
\altaffiltext{6}{Department of Physics, University of California at Santa Cruz, Santa Cruz, CA 95064 USA}

\begin{abstract}
We find that infalling dark matter halos (i.e., the progenitors of satellite halos) begin losing mass well outside the virial radius of their eventual host halos.  The peak mass occurs at a range of clustercentric distances, with median and 68$^\mathrm{th}$ percentile range of $1.8^{+2.3}_{-1.0} R_\mathrm{vir,host}$ for progenitors of $z=0$ satellites. The peak circular velocity for infalling halos occurs at significantly larger distances ($3.7^{+3.3}_{-2.2} R_\mathrm{vir,host}$ at $z=0$).  This difference arises because different physical processes set peak circular velocity (typically, $\sim$ 1:5 and larger mergers which cause transient circular velocity spikes) and peak mass (typically, smooth accretion) for infalling halos.  We find that infalling halos also stop having significant mergers well before they enter the virial radius of their eventual hosts.  Mergers larger than a 1:40 ratio in halo mass end for infalling halos at similar clustercentric distances ($\sim 1.9 R_\mathrm{vir,host}$) as the end of overall mass accretion.  However, mergers larger than 1:3 typically end for infalling halos at more than 4 virial radial away from their eventual hosts. This limits the ability of mergers to affect quenching and morphology changes in clusters.  We also note that the transient spikes which set peak circular velocity may lead to issues with abundance matching on that parameter, including unphysical galaxy stellar mass growth profiles near clusters; we propose a simple observational test to check if a better halo proxy for galaxy stellar mass exists.
\end{abstract}
\keywords{dark matter --- galaxies: abundances --- galaxies:
  evolution --- methods: N-body simulations}

\section{Introduction}

\label{s:intro}

In the Lambda Cold Dark Matter ($\Lambda$CDM) paradigm, galaxies form in the centers of dark matter halos, and the growth of both the galaxy stellar mass and halo mass are strongly correlated \citep[see][for recent constraints, and references therein]{cw-08,Behroozi10,Behroozi13, BWC12, moster-09, Moster12, Leauthaud12, Yang11, Wang12, Bethermin13}.  Dark matter halos may be classified into two types depending on whether the center is contained within a larger halo (satellite halos or subhalos), or whether it is not (host halos).  Similarly, galaxies may be divided into centrals and satellites, depending on whether they reside in the center of a host halo or in the center of a smaller, satellite halo.

It is well-known both that satellite dark matter halos rapidly lose mass due to tidal forces \citep[e.g.,][]{Tormen98b,Kravtsov04b,Knebe06} and that satellite galaxies quench (i.e., stop forming stars) after accretion onto a cluster \citep[see recently, e.g.,][and references therein]{Yang07,Kimm09,Prescott11,Wetzel11,Woo13}.  Because halo mass accretion is strongly connected to gas accretion \citep{vdVoort11}, it is not surprising that a diminishing fuel supply correlates with the end of star formation.  However, it has been known for over a decade that enhanced galaxy quenching exists well past the virial radius of clusters \citep{Balogh00,Verdugo08,Braglia09,Wetzel13}.  It is also not clear to what extent quenching is triggered by a single external event (such as a merger), or whether it happens due to more gradual harassment \citep{Croton06,Somerville08,Tecce10,Book10,Lu11,Kimm11,Wetzel11}.  Recently, \cite{Hahn09}, \cite{Reddick12}, and \cite{Bahe13} noted that \textit{halos} can also lose mass beyond the virial radii of clusters.  Less attention has been given to the distances where infalling halos experience their last mergers.  In this paper, we investigate the 
range of clustercentric distances at which subhalos start to be dynamically influenced by their hosts, evaluating the
radii characterizing the beginning of mass loss and the last merger for a range of mass ratios.  We discuss the implications of these results for satellite quenching and abundance matching models.

Maximum circular velocity ($v_\mathrm{max} = \max(\sqrt{GM(<R)/R})$), in addition to halo mass, is highly correlated with galaxy stellar mass \citep{Colin99,conroy:06,wetzel-09,TG11,Bolshoi,Reddick12}.  $\vmax$ may in fact be a better halo proxy for stellar mass than is halo mass for host halos, because $\vmax$ directly measures the gravitational potential close to the galaxy center \citep{Colin99,conroy:06,Bolshoi,TG11}.  
\cite{conroy:06} suggested using $\vmax$ measured at the epoch of accretion for satellite halos, because 
$\vmax$ also decreases after infalling halos become satellites, although it does so significantly less than halo mass.
By abundance matching galaxies rank-ordered by luminosity to halos rank-ordered by $\vmax$ (measured at accretion for satellites) in equal volumes, \cite{conroy:06} found that the halos matched the galaxy clustering statistics at a range of redshifts.  However, \cite{Reddick12} found that abundance matching galaxies to halos by \textit{peak} $\vmax$ (i.e., the highest $\vmax$ ever reached in a halo's accretion history) may give a better match to more recent clustering constraints.  In this paper, we calculate clustercentric distances at which halos reach peak $\vmax$ to determine if there is a physical reason for peak $\vmax$ to be a better stellar mass proxy than $\vmax$ at accretion. We note that here we use the term ``clustercentric'' distance to refer to the distance to the halo that a satellite will eventually fall into, regardless of the mass of the host halo.

For all results, we use dark matter-only simulations; however, we test with several different combinations of simulation codes, halo finders, and cosmological parameters.  We provide details on the simulations, halo finding, and merger trees in \S \ref{s:methods}.  Quantitative results for where infalling halos reach their peak masses and circular velocities, as well as have their last mergers, are presented in \S \ref{s:rpeak}.  Finally, we discuss how the results impact satellite galaxy quenching, satellite morphologies, and abundance matching in \S \ref{s:discussion}, and conclude in \S \ref{s:conclusions}.  Our main results in this paper assume a flat, $\Lambda$CDM cosmology with parameters $\Omega_M = 0.27$, $h=0.7$, $n_s = 0.95$, and $\sigma_8 = 0.82$.  Halo masses are defined using the virial spherical overdensity criterion of \cite{mvir_conv}.

\begin{figure*}
\plotgrace{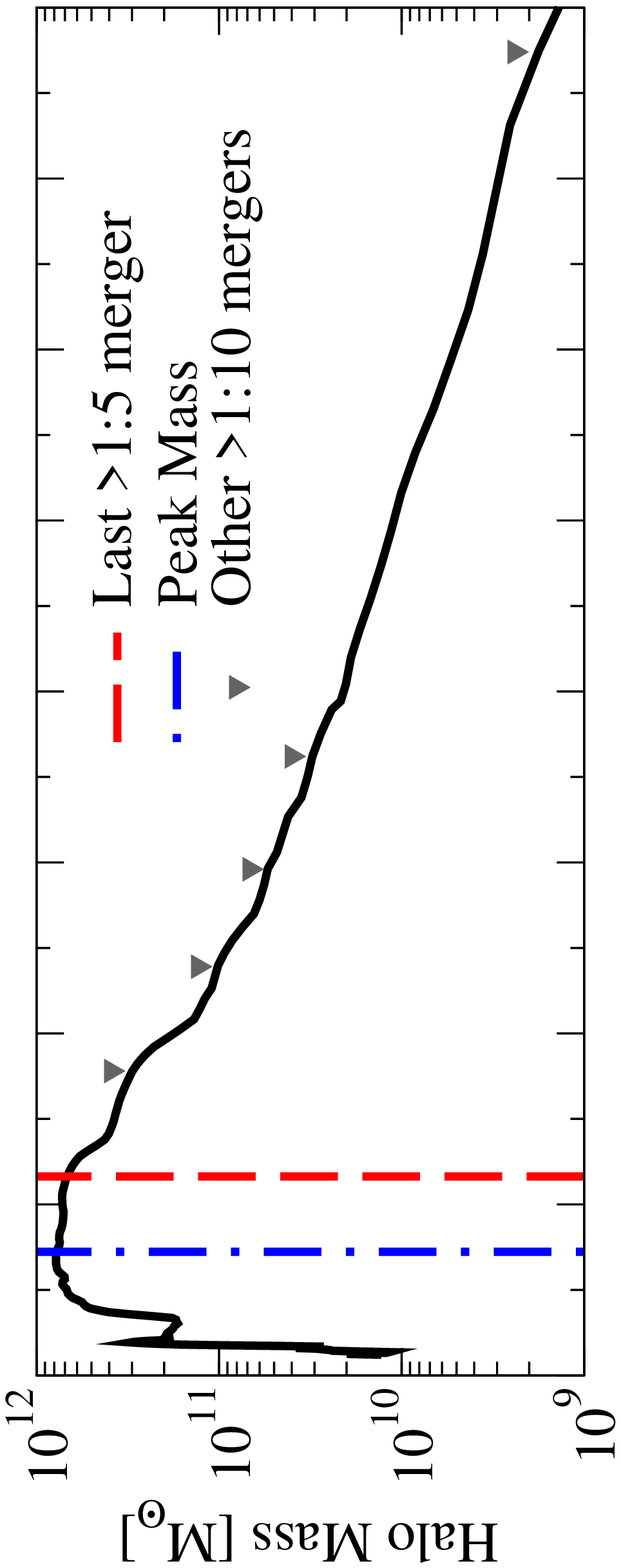} \plotgrace{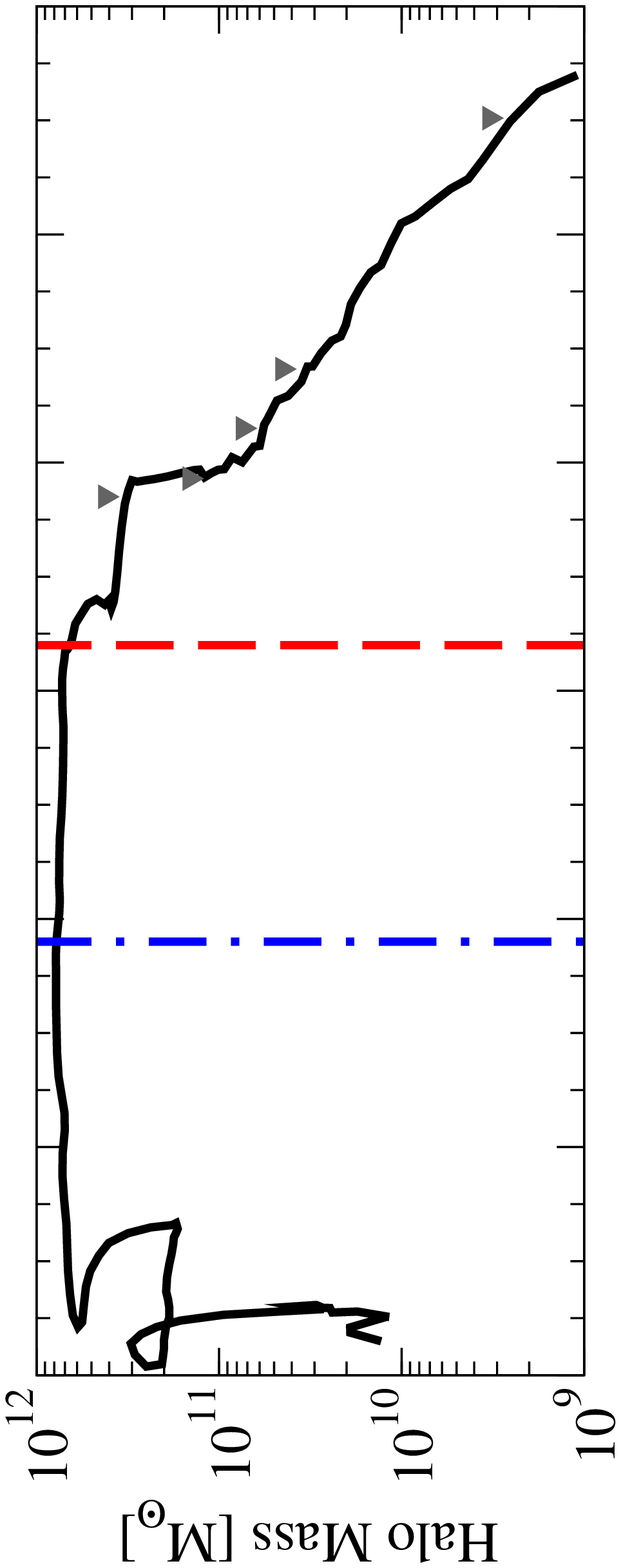}\\[-31ex]
\plotgrace{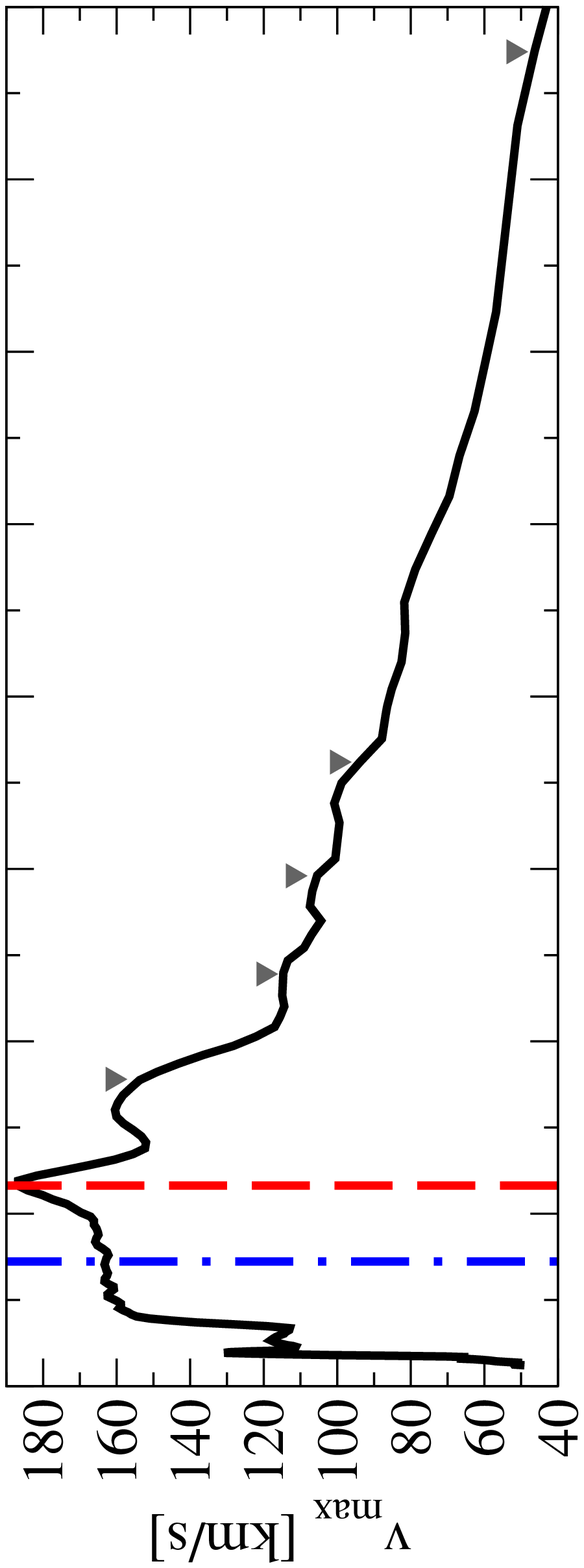} \plotgrace{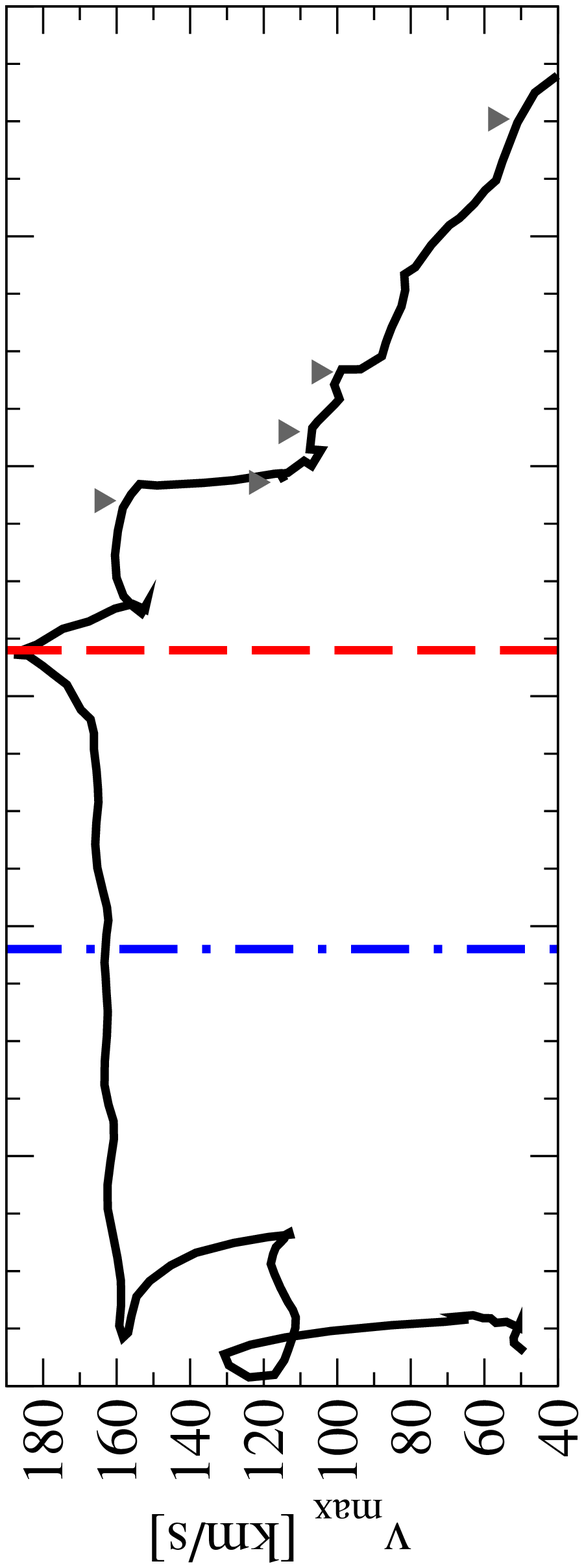}\\[-31ex]
\plotgrace{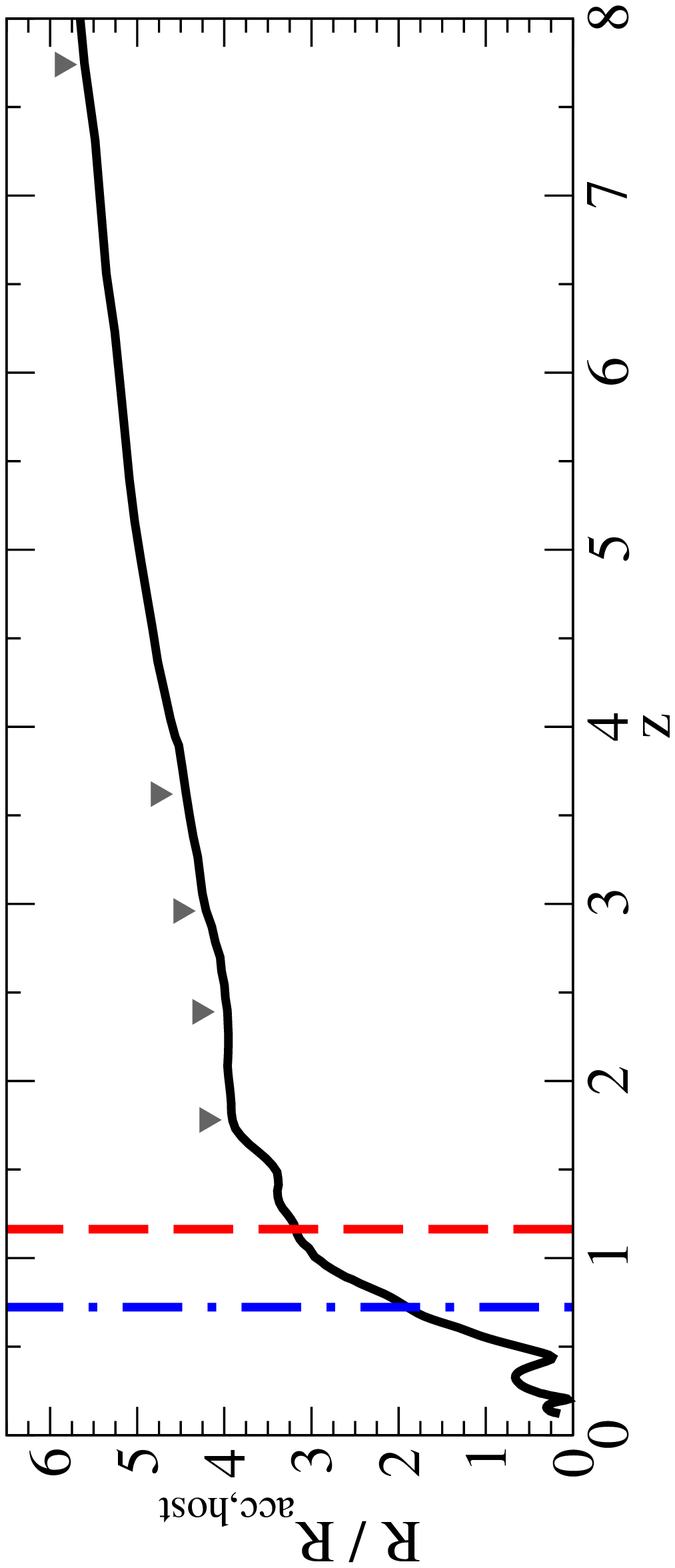} \plotgrace{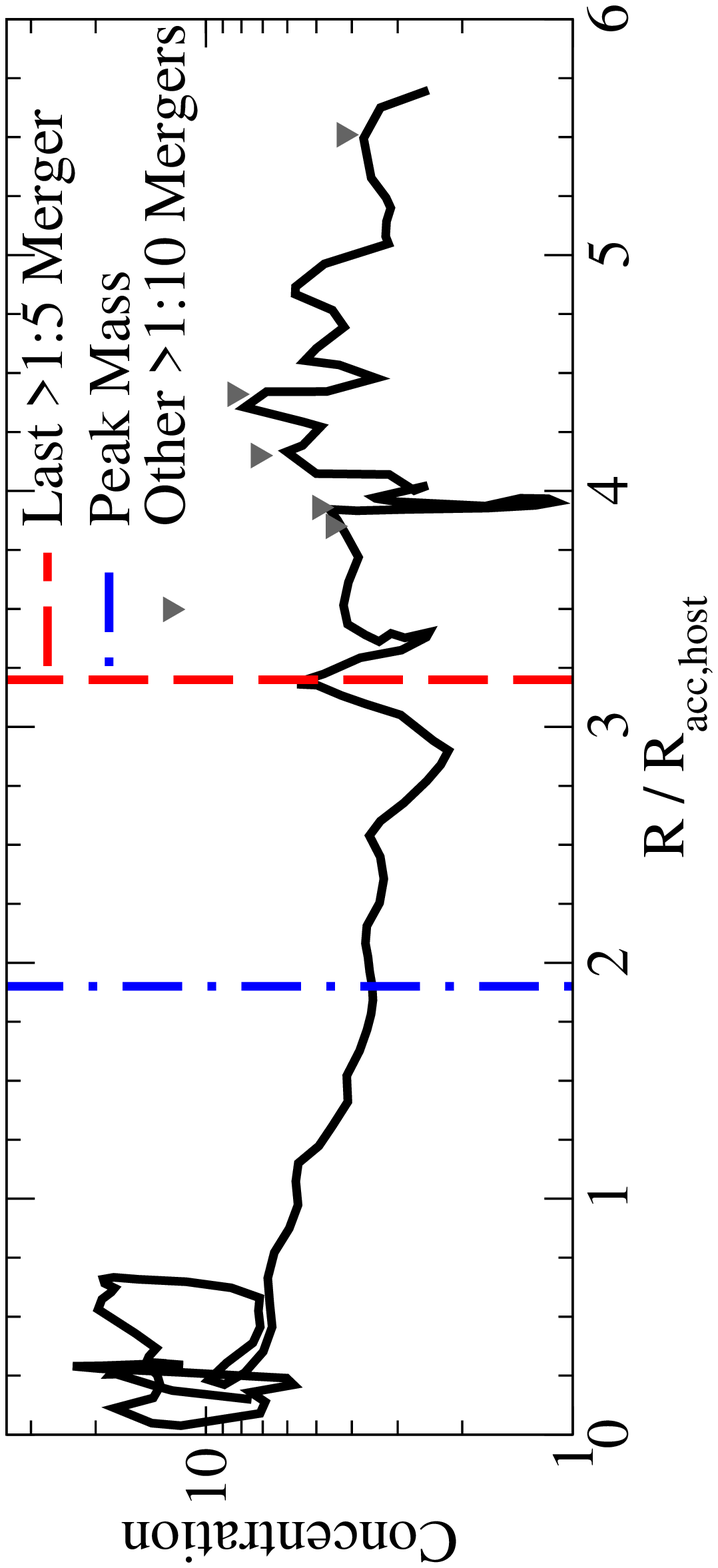}\\[-20ex]
\caption{\textbf{Left panels} show halo mass and $\vmax$ histories for a typical infalling halo; \textbf{right panels} show halo mass, $\vmax$, and concentration histories for the same halo as a function of the distance to its eventual host ($R$), in units of the host virial radius at accretion ($R_\mathrm{acc,host}$).  Peak mass and peak $\vmax$ occur at different times and distances.  Peak $\vmax$ occurs first, when a merging satellite halo on its first pass through the infalling halo's center creates a temporary spike in concentration as well as $\vmax$.  This event is marked by a \textbf{red dashed line} in all plots.  Peak mass occurs second, and is due to slow, steady accretion; this is marked by the \textbf{blue dot-dashed line} in all plots.  However, the infalling halo reaches peak mass well outside of the virial radius of its eventual host ($\sim 2 R_\mathrm{acc,host}$, in this example).  Steady tidal stripping leads to a slow decline in both mass and $\vmax$; rapid mass loss does not occur until the infalling halo is well inside its host.  Other $>$1:10 mergers are marked by \textbf{grey triangles}, specifically at the time of first turnaround of the merging satellite. These mergers all result in spikes in concentration and $\vmax$, but the effect is more pronounced for more massive mergers.  All quantities are calculated using the \textsc{Rockstar} halo finder on the Bolshoi simulation.}
\label{f:example}
\end{figure*}

\section{Methods}

\label{s:methods}

We here briefly overview the simulations (\S \ref{s:sims}), halo finders (\S \ref{s:hf}), and merger tree code (\S \ref{s:mt}) which we have used.

\subsection{Simulations}
\label{s:sims}
We make use of the \textit{Bolshoi} simulation \citep{Bolshoi}, which follows 2048$^3$ dark matter particles in a (250 $h^{-1}$ Mpc)$^3$ volume using the \textsc{art} code \citep{kravtsov_etal:97}.  Its excellent mass ($1.94\times 10^{8}\Msun$ per particle) and force resolution (1 $h^{-1}$ kpc) make it ideal for studying satellite halos down to $10^{10.5}\Msun$ in host halos as large as $10^{15}\Msun$.  The assumed $\Lambda$CDM cosmology is close to the WMAP9 best-fit cosmology \citep{WMAP9}, with parameters $\Omega_M = 0.27$, $\Omega_\Lambda = 0.73$, $h=0.7$, $n_s = 0.95$, and $\sigma_8 = 0.82$.

We also make use of the \textit{Consuelo} simulation (McBride et al., in prep.; see also \citealt{Rockstar,BehrooziTree,Wu13}) to verify that our results are not sensitive to the above choice of simulation code or cosmology.  Consuelo follows 1400$^3$ dark matter particles in a larger (420 $h^{-1}$ Mpc)$^3$ volume using the \textsc{gadget} code \citep{Springel05}.  Its force and mass resolution are 8 $h^{-1}$ kpc and $2.7\times 10^{9}\Msun$, respectively; the assumed flat $\Lambda$CDM cosmology has parameters $\Omega_M = 0.25$, $\Omega_\Lambda = 0.75$, $h=0.7$, $n_s = 1.0$, and $\sigma_8 = 0.8$.

\subsection{Halo Finders}
\label{s:hf}
Both simulations were analyzed using the \textsc{Rockstar} halo finder \citep{Rockstar}.  This halo finder is a fully phase-space temporal (7D) algorithm featuring excellent recovery of major merger and satellite halo properties \citep{Knebe11,Onions12,Onions13,BehrooziTree}. The method uses adaptively shrinking phase-space linking lengths to find density peaks in phase space.  Particles which share a common closest density peak in phase space are grouped into a single halo or satellite halo, whose position is the location of the density peak.  Then, particles with positive total energy are removed using a tree-code calculation and full halo properties (including velocities, virial masses, and maximum circular velocities) are calculated.  When multiple simulation timesteps are available, they are used to ensure continuity of host halo / satellite halo relationships in cases (such as major mergers) where they may be ambiguous.

To verify that our results are not sensitive to the halo finder used, we also have run the \textsc{BDM} halo finder \citep{Klypin99,Riebe12} on the Bolshoi simulation.  This halo finder is a position-space (3D) algorithm which calculates densities for each particle via a top-hat filter over the nearest 20 particles.    Around each density maxima, \textsc{BDM} grows spherical shells until the enclosed mass corresponds to a specified density threshold $\Delta$ (in this case, $\Delta_\mathrm{vir}$; \citealt{mvir_conv}).  By definition, this means that all halo centers will correspond to density maxima.  These spherical regions are converted to halos in order of deepest central gravitational potential; satellite mass profiles are truncated if they exceed the distance to the nearest larger halo center by a tunable overshoot factor (1.1-1.5).  For satellite halos only, iterative unbinding using spherically-averaged potentials is performed before determining halo properties.\footnote{Earlier versions of \textsc{BDM} performed iterative unbinding on all halos, not just satellites.}

\subsection{Merger Trees}
\label{s:mt}
Merger trees for all simulations were generated using the \textsc{Consistent Trees} algorithm of \cite{BehrooziTree}.  This approach simulates the gravitational motion of halos given their positions, velocities, and mass profiles as returned by the halo finder.  From information in a halo catalog at a given simulation snapshot, the expected positions and velocities of halos at an earlier snapshot may be calculated.  In cases where there is an obvious inconsistency between the expected halo positions and the actual ones as returned by the halo finder, the halo catalog can be repaired by substituting the expected halo properties.  This process repairs defects such as missed satellite halos (e.g., satellite halos which pass too close to the center of a larger halo to be detected) and spurious mass changes (e.g., satellite halos which suddenly increase in mass due to temporary misassignment of particles from the host halo).  This process is very important to ensure accurate mass accretion histories for satellite and host halos; full details of the algorithm as well as tests of the approach applied to both Bolshoi and Consuelo simulations may be found in \cite{BehrooziTree}.

\section{Results}
\label{s:rpeak}

In this section, we trace merger histories of satellite dark matter halos to determine the clustercentric distances where they reach peak $\vmax$ and mass, as well as where they had their last mergers.   For infalling halos, their peak mass is expected to occur around $\sqrt[3]{3} \sim 1.4$ times the radius of their eventual host ($R_\mathrm{vir,host}$) \citep{Hahn09}.  This is the distance at which stable orbits at the radius of the infalling halo no longer exist; i.e., it is where the Hill (also known as ``Roche'' or ``tidal'') radius coincides with the virial radius of the infalling halo.\footnote{More generally, this ratio applies for any two spheres with the same average density.  This means that for \textit{any} given spherical overdensity $\Delta$, the peak mass $M_\Delta$ for an infalling halo will happen near $\sqrt[3]{3}$ times the eventual host halo's radius measured at the same overdensity (i.e., $R_{\Delta,\mathrm{host}}$).}  To better compare clustercentric distances across different host masses, we normalize all distances by the virial radius of the eventual host at the time of last accretion ($R_{\rm acc, host} = R_{\rm vir, host} (z_{\rm acc})$).  Our results do not change appreciably if we normalize by the host's virial radius at a different time (e.g., at the time the infalling halo reached peak mass or $\vmax$).

We first present mass, $\vmax$, and concentration histories for a typical infalling halo in \S \ref{s:example} in order to motivate the analysis that follows.  We discuss results for the radius at which satellite halos reach peak $\vmax$ ($R_\mathrm{peak,vmax}$) in \S \ref{s:rpeak_vmax} and identify the connection to the last $>$1:5 merger in \S \ref{s:last_mm}.  We then compare $R_\mathrm{peak,vmax}$ to the radius of peak mass ($R_\mathrm{peak,mass}$) in \S \ref{s:rmass}.  We examine the percentage of satellites which reach peak mass and $\vmax$ after ``backsplash'' (i.e., after passing temporarily through the virial radius of a larger halo) in \S \ref{s:backsplash}.  Finally, we compare results across different halo finders and simulations in \S \ref{s:hf_sims}.

\subsection{Example}
\label{s:example}

We show the halo mass, $\vmax$, merging, and concentration histories for a typical infalling halo in Fig.\ \ref{f:example}.  As time proceeds, the halo has monotonic growth in mass, but its $\vmax$ growth is marked by temporary spikes.  These spikes in $\vmax$ often correspond to mergers (grey triangles in Fig.\ \ref{f:example}) as well as to spikes in concentration, suggesting that a merging satellite passing by the halo's center is causing a temporary boost in central density.  The $\vmax$ peaks are generally larger for larger merger ratios, such as the major 1:3 merger at $z = 1.78$, as compared to the mostly $\sim$ 1:10 events at higher redshifts for this halo.  This halo's peak $\vmax$ is set during a $\sim$1:5 merger at $z=1.17$, which occurs at $\sim$3 times the virial radius of its eventual host.  After this peak, $\vmax$ declines immediately by 12\%.  For this halo, peak $\vmax$ does not correspond to peak mass.  Instead, its mass continues to grow through smooth accretion until the halo reaches $\sim2$ times the virial radius of its eventual host.  At this radius, tidal forces from the host are strong enough to halt accretion, even though severe stripping does not occur until after the infalling halo passes through the virial radius of its host.

As shown in later sections, several aspects of this example apply to the halo population as a whole.  Peak $\vmax$ tends to occur well outside the virial radius of the eventual host (\S \ref{s:rpeak_vmax}), and is very often coincident with a $>$1:5 merging event (\S \ref{s:last_mm}).  Peak mass is often set instead by a different process---the tidal truncation of smooth accretion---and therefore occurs much closer to the virial radius of the eventual host halo (\S \ref{s:rmass}).  We demonstrate each of these findings on large statistical samples in the next sections.

\subsection{The Radius of Peak Satellite Halo $\vmax$}
\label{s:rpeak_vmax}

\begin{figure}
\plotgrace{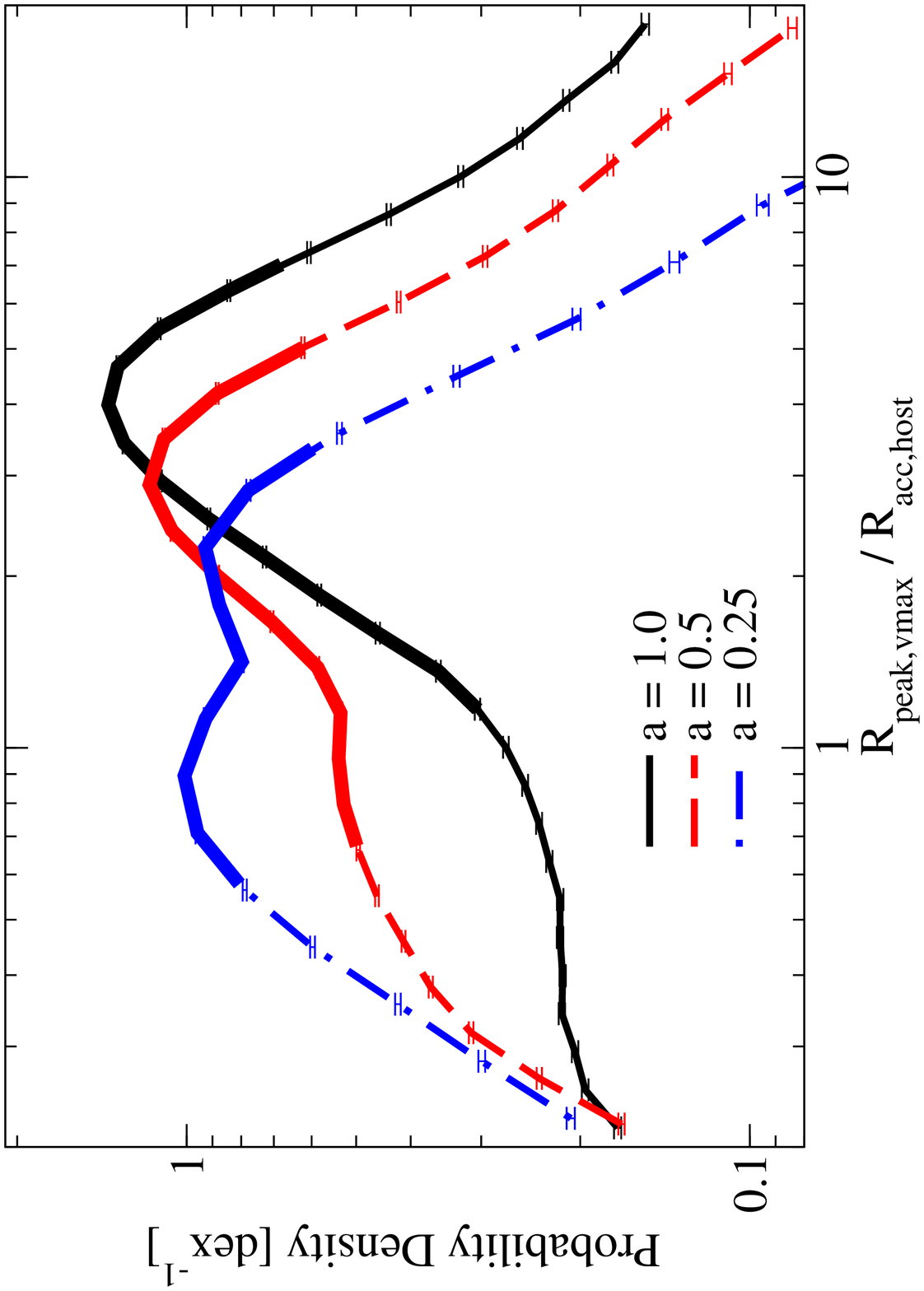}\\[-5ex]
\plotgrace{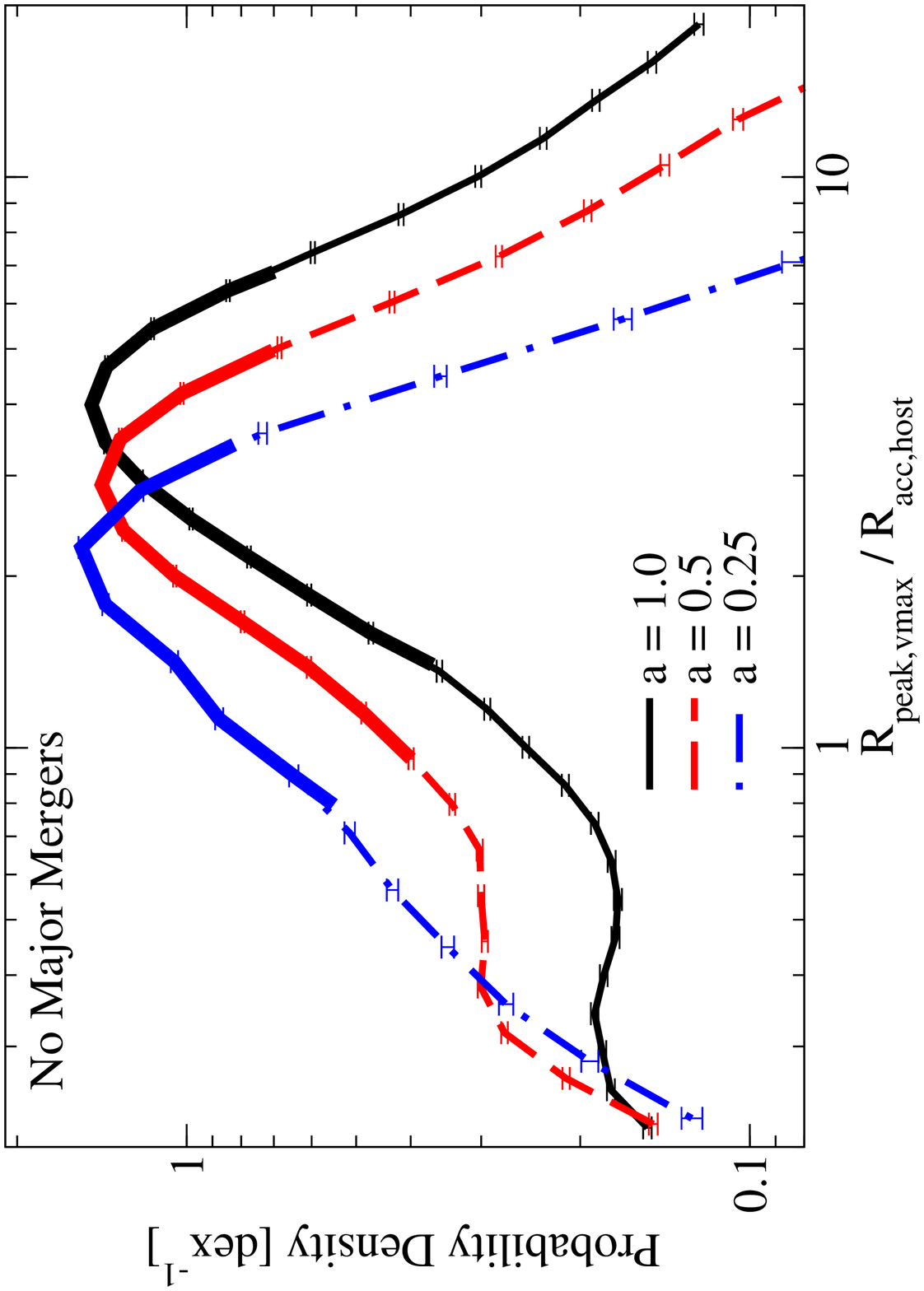}
\caption{\textbf{Top panel}: the distribution of $R_\mathrm{peak,vmax} / R_\mathrm{acc,host}$ for subhalos at several redshifts in the Bolshoi simulation with the \textsc{Rockstar} halo finder.  $R_\mathrm{peak,vmax}$ is the distance from the host halo at which the satellite halo progenitor had its maximum $\vmax$; $R_\mathrm{acc,host}$ is the virial radius of the host halo at the epoch of the satellite halo's accretion.  \textbf{Double-width lines} correspond to the middle 68\% of the probability distributions.  At higher redshifts, satellites which are major mergers make up a larger percentage of the satellite population studied; major mergers tend to reach their peak $\vmax$ within the radius of their host halo.  \textbf{Bottom panel}: same as middle panel, excluding major ($>$1:3) mergers.}
\label{f:rpeak2}
\end{figure}

\begin{figure}
\plotgrace{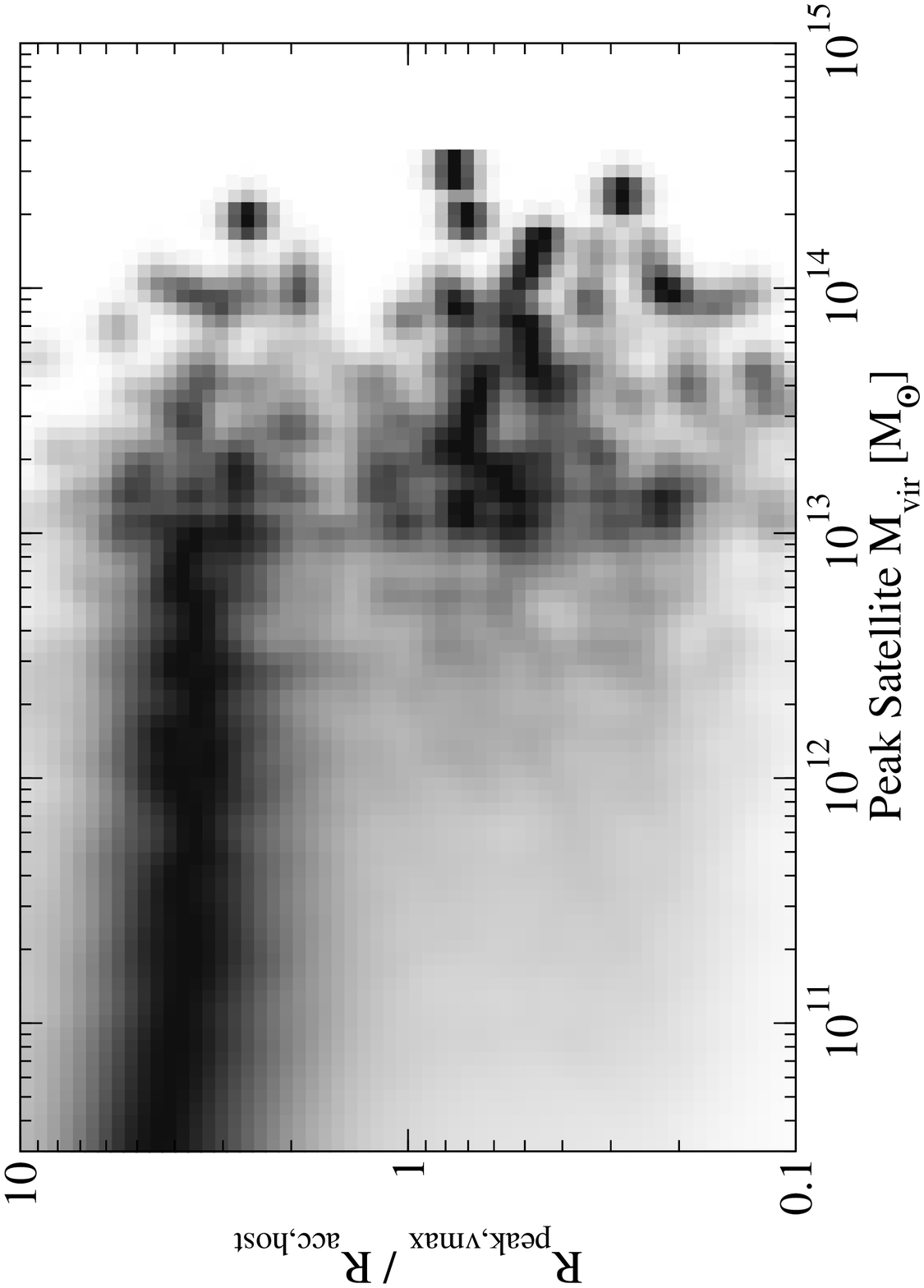}\\[-5ex]
\plotgrace{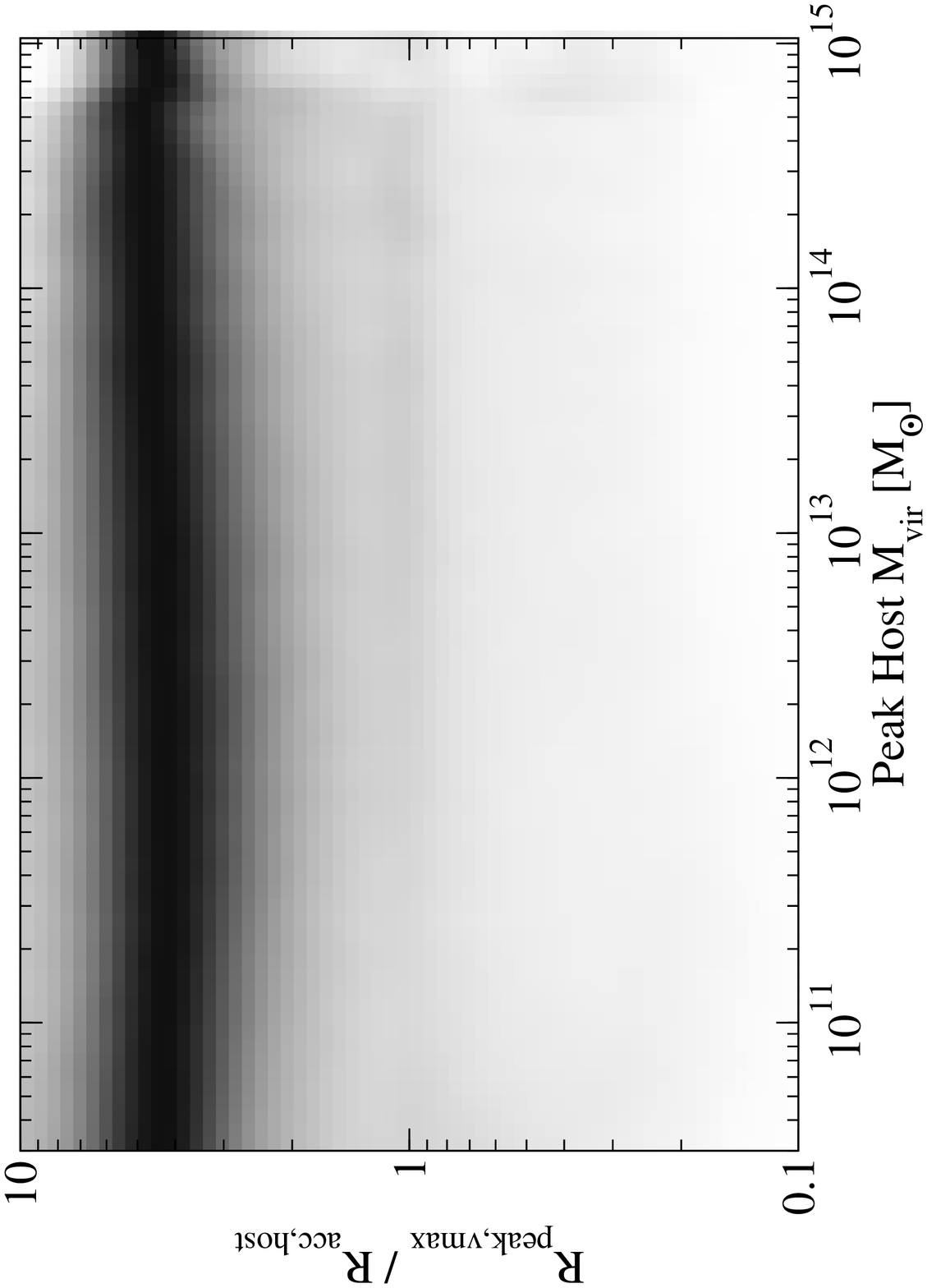}
\caption{Conditional density plot of $R_\mathrm{peak,vmax} / R_\mathrm{acc,host}$ as a function of satellite halo mass (\textbf{top panel}) and host mass (\textbf{bottom panel}) at $z=0$ in Bolshoi with the \textsc{Rockstar} halo finder.  As in Fig.\ \ref{f:rpeak2}, $R_\mathrm{peak,vmax}$ is the distance from the final host at which the satellite halo progenitor had its maximum $\vmax$; $R_\mathrm{acc,host}$ is the radius of the host halo at the epoch of the satellite halo's accretion.  The region below a ratio of unity is largely populated by major ($>$1:3) mergers and halos which undergo satellite-satellite mergers (see \S \ref{s:rpeak_vmax}).  Otherwise, no strong mass trends are evident.  A smoothing kernel with a FWHM of 0.12 dex has been applied to both plots; individual high-mass satellite halos therefore appear as round blobs.}
\label{f:rpeak}
\end{figure}

Figure \ref{f:rpeak2} demonstrates that most satellites reached their peak $\vmax$ well outside the host halo in which they currently reside.  The median $R_\mathrm{peak,vmax}$ for $z=0$ satellites is at 3.7 times $R_\mathrm{acc,host}$, as shown in the upper panel of Fig.\ \ref{f:rpeak2}; the $68^\mathrm{th}$-percentile range is extremely large: $1.5-7 R_\mathrm{acc,host}$.  At higher redshifts, infalling halos reach peak $\vmax$ much closer to their final hosts; we defer the explanation of this to \S \ref{s:last_mm}.  The halos which reach peak $\vmax$ after becoming satellites are those which experience chance satellite-satellite mergers, as well as major ($>$1:3) mergers.

Major mergers are unique because the merging halos can continue to accrete weakly-bound material even after they pass within the host halo's virial radius.  These mergers account for a satellite mass dependence in $R_\mathrm{peak,vmax}$ (Fig.\ \ref{f:rpeak}, top panel).  Major mergers are more common for satellite halo masses above $10^{13}\Msun$; this is because of the exponential decline in the host halo mass function towards higher masses---i.e., the rarity of host halos for which a $>10^{13}\Msun$ satellite would \textit{not} be considered a major merger.  If the distribution of $R_\mathrm{peak,vmax}$ is instead plotted as a function of host halo mass (Figure \ref{f:rpeak}, lower panel), the small percentage of major mergers in the overall merger ratio spectrum means that no strong mass trends are apparent.

We find that at higher redshifts, more surviving satellites are major mergers, meaning that a larger proportion of halos reach peak $\vmax$ within the virial radius of their host halos.  Excluding major mergers, the probability distributions for peak $\vmax$ are more similar in shape (Fig.\ \ref{f:rpeak2}, bottom panel).

\begin{figure}
\plotgrace{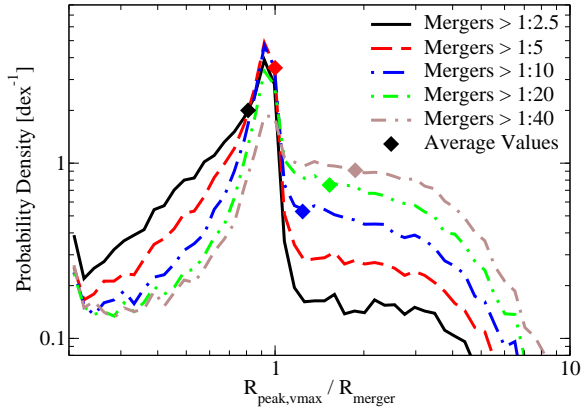}
\caption{Histograms of $R_\mathrm{peak,vmax} / R_\mathrm{merger}$ for satellites at $z=0$; several different merger mass ratio thresholds are shown.  $R_\mathrm{peak,vmax}$ greater than $R_\mathrm{merger}$ almost always implies that the merger occurred after peak $\vmax$ was reached; $R_\mathrm{peak,vmax}$ less than $R_\mathrm{merger}$ implies the opposite.  While major mergers almost always lead to a new peak in $\vmax$ (see, e.g., the line corresponding to $>$1:2.5 mergers), smaller mergers can sometimes cause $\vmax$ to peak as well.  The average of $R_\mathrm{peak,vmax} / R_\mathrm{merger}$ (marked by \textbf{diamonds} for each merger threshold) is closest to 1 when $R_\mathrm{merger}$ refers to the last $>$1:5 merger (see also \S \ref{s:last_mm}).  Less massive mergers are much less likely to set peak $\vmax$, and so continue to occur after peak $\vmax$ is set (i.e., $R_\mathrm{peak,vmax}> R_\mathrm{merger}$).  More massive mergers almost always set peak $\vmax$, but later 1:5 mergers have some chance of setting peak $\vmax$, resulting in $R_\mathrm{peak,vmax}< R_\mathrm{merger}$.  This suggests that a $>$1:5 merger event is the best proxy for peak $\vmax$ among the options considered here.  To avoid mass resolution limits, these values were calculated only for halos larger than $4\times 10^{11}\Msun$ for Bolshoi with the \textsc{Rockstar} halo finder.}
\label{f:merger_hist}
\end{figure}

\begin{figure}
\plotgrace{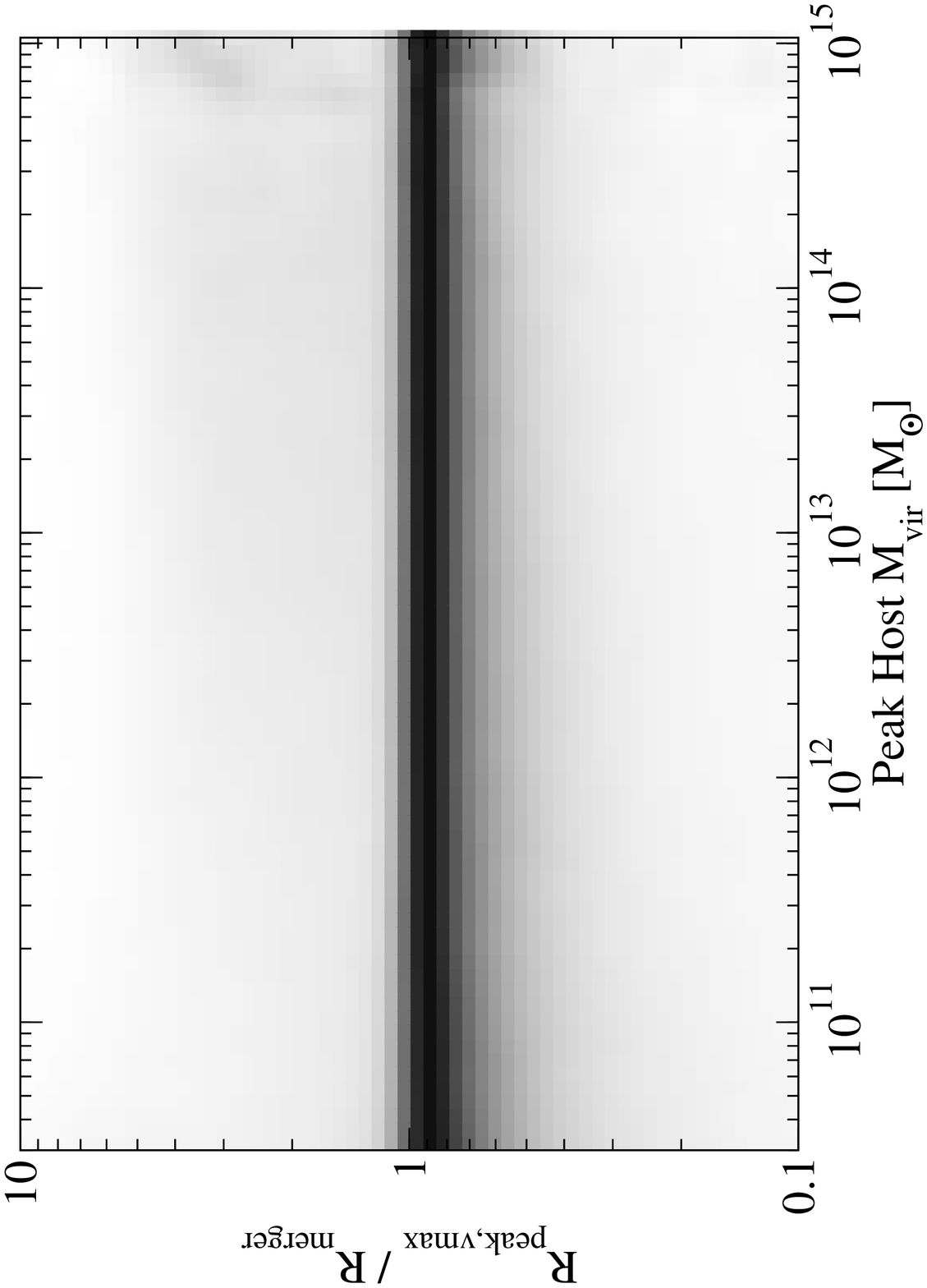}
\caption{Conditional density plot of $R_\mathrm{peak,vmax} / R_\mathrm{merger}$ as a function of host mass for satellite halos at $z=0$ in the Bolshoi simulation.  $R_\mathrm{merger}$ is the radius at which the satellite halo had its last minor (1:5) merger.}
\label{f:mm}
%\plotgrace{graphs/vmax_conc}
%\caption{$\vmax$ as a function of virial mass at $z=0$ for a \citep{NFW97} halo profile.  When a merging satellite makes its first pass close to the halo center, the host halo concentration will temporarily increase, leading to a spike in $\vmax$.}
%\label{f:vmax}
\plotgrace{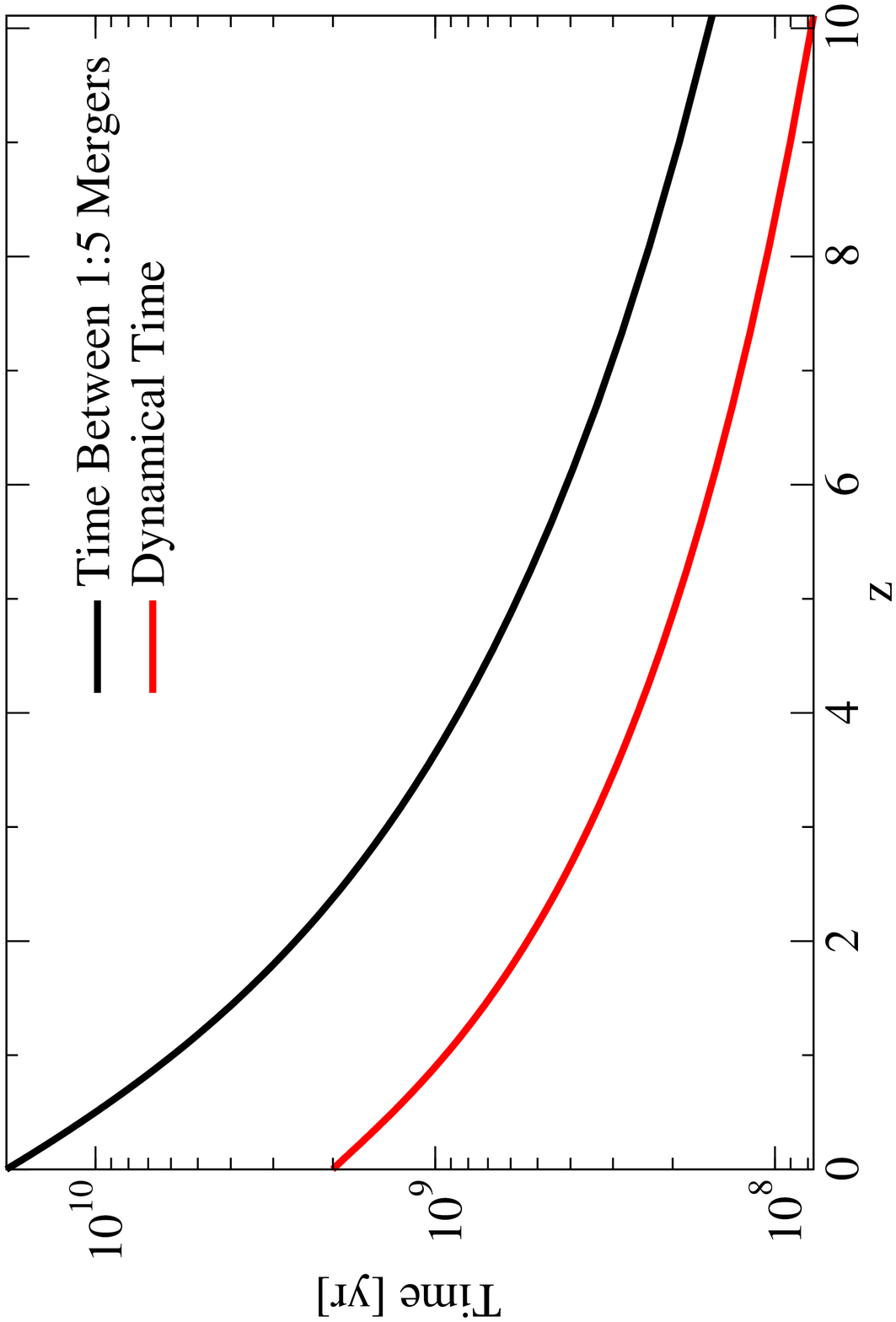}
\caption{Comparison between the dynamical time ($(G \rho_{\rm vir})^{-1/2}$) and the average time between 1:5 or larger mergers, from \cite{BWC12}.  As the dynamical time is much shorter than the merger time at low redshifts, infalling halos will tend to travel longer distances between experiencing mergers; this qualitatively explains the high values of $R_\mathrm{peak,vmax}/R_\mathrm{acc,host}$ seen in Figs.\ \ref{f:rpeak2} and \ref{f:rpeak}.}
\label{f:time_comp}
\end{figure}

\subsection{The Relationship Between Peak $\vmax$ and the Last Minor Merger}

\label{s:last_mm}

The fact that peak $\vmax$ for satellite progenitors occurs at such a large distance from their hosts cannot be explained by tidal stripping.  Indeed, as shown in \S \ref{s:example} and \S \ref{s:rmass}, infalling halos can continue accreting mass well after peak $\vmax$ is reached.  Instead, we find that peak $\vmax$ is highly correlated with the last $>$1:5 merger, as shown in Figs.\ \ref{f:merger_hist} and \ref{f:mm}.  This is because $\vmax$ is generally set by the average density near the halo center.  If a halo experiences a merger, the merger will temporarily boost the central density of the halo as it makes its first pass near the halo's center.  This will in turn temporarily boost both $\vmax$ and the concentration of the halo, as seen in Fig.\ \ref{f:example} \citep[see also][]{Ludlow12}.  Once the merging halo orbits and tidally dissipates, it raises the velocity dispersion of its host, which lowers both $\vmax$ and concentration again.  For this reason, minor and larger mergers introduce a temporary spike in $\vmax$ for the infalling halo, typically about 14\% (Appendix \ref{a:timing}).  Because $\vmax$ is roughly proportional to the cube root of halo mass, the mass of the halo would have to grow by 42\% before the halo would again reach the peak $\vmax$ set by the merging event.  The fact that mergers can easily set peak $\vmax$ applies equally well for host halos, as shown in Appendix \ref{a:timing}.

The probability for a merger event to set peak $\vmax$ depends strongly on the merger ratio: mergers with higher mass ratios are more likely to pass closer to the center, and will also lead to larger density increases.  However, the infrequency of major mergers means that it is possible for smaller, more frequent minor mergers to set a new peak $\vmax$ after the last major merger event.  We find that the radius at which the last $>$1:5 merger occurred is on average the same as $R_\mathrm{peak,vmax}$ (Fig.\ \ref{f:merger_hist}).

We can then understand why peak $\vmax$ occurs at large clustercentric distances by comparing the gravitational timescale for free-fall (i.e., the dynamical timescale) to the timescale for $>$1:5 mergers.  As shown in Fig.\ \ref{f:time_comp}, the merger timescale is much longer than the dynamical timescale at $z=0$, meaning that infalling halos will travel for significant distances (relative to the host's virial radius) between mergers.  This ratio becomes less at higher redshifts, meaning that the travel distance will be shorter relative to the host's radius.  While a direct quantitative comparison cannot be made---e.g., infalling halos do not start from rest at the time of their last merger, and tidal forces will reduce the merger rate close to the host halo \citep{Binney08}---it is instructive to calculate the expected radial distances for the last $>$1:5 merger based on the ratio of the time between mergers to the dynamical time.  For free-falling orbits, the travel distance is proportional to the travel time to the two-thirds power, so we can very roughly estimate
\begin{equation}
\frac{R_\mathrm{merger}}{R_\mathrm{acc,host}} \sim \left(\frac{T_\mathrm{merger}}{T_\mathrm{dyn}}\right)^\frac{2}{3}
\end{equation}
where $T_\mathrm{merger}$ is the average time between mergers, $T_\mathrm{dyn}$ is the dynamical time $((G \rho_{\rm vir})^{-1/2})$, and $\rho_\mathrm{vir}$ is the virial overdensity.    The expected distances are then 4.4, 3.4, and 2.5 $R_\mathrm{acc,host}$ at $z=0$, $1$, and $3$, respectively.  Despite the crudeness of this calculation, these values capture the redshift scaling and normalization of the average $R_\mathrm{peak,vmax}$ found in our simulations when major mergers are excluded (e.g., Fig.\ \ref{f:rpeak2}, bottom panel).

\begin{figure}
\plotgrace{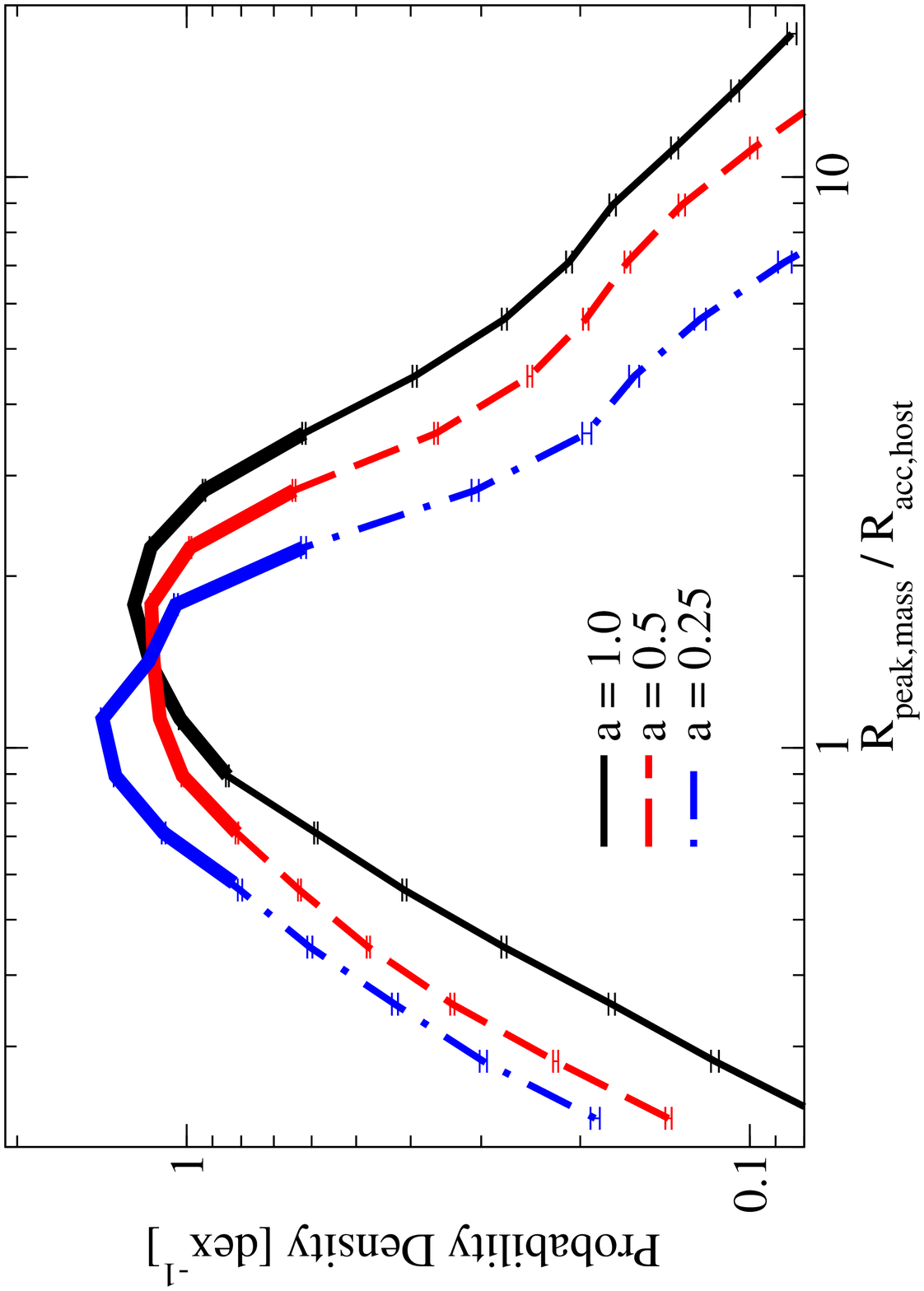}
\plotgrace{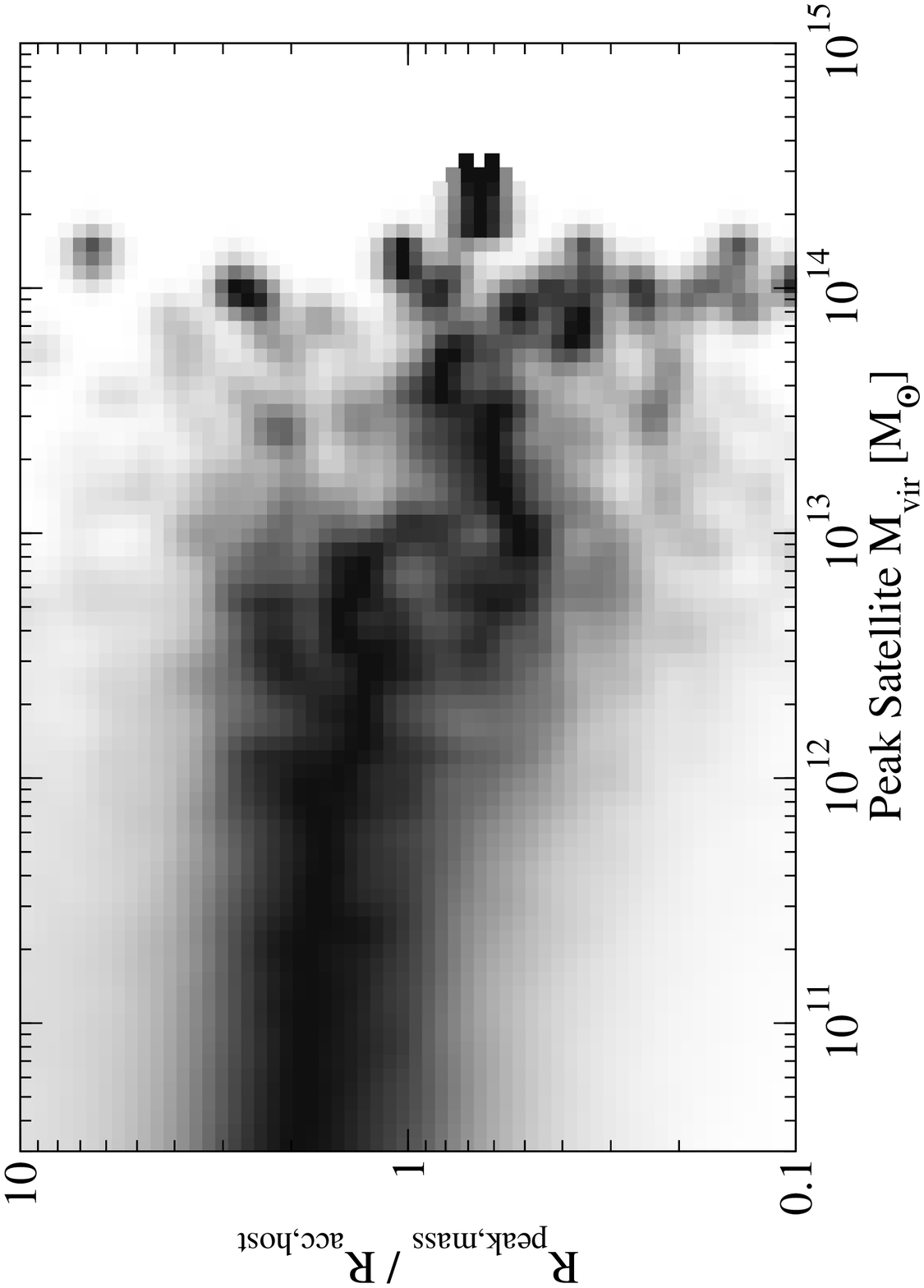}
\plotgrace{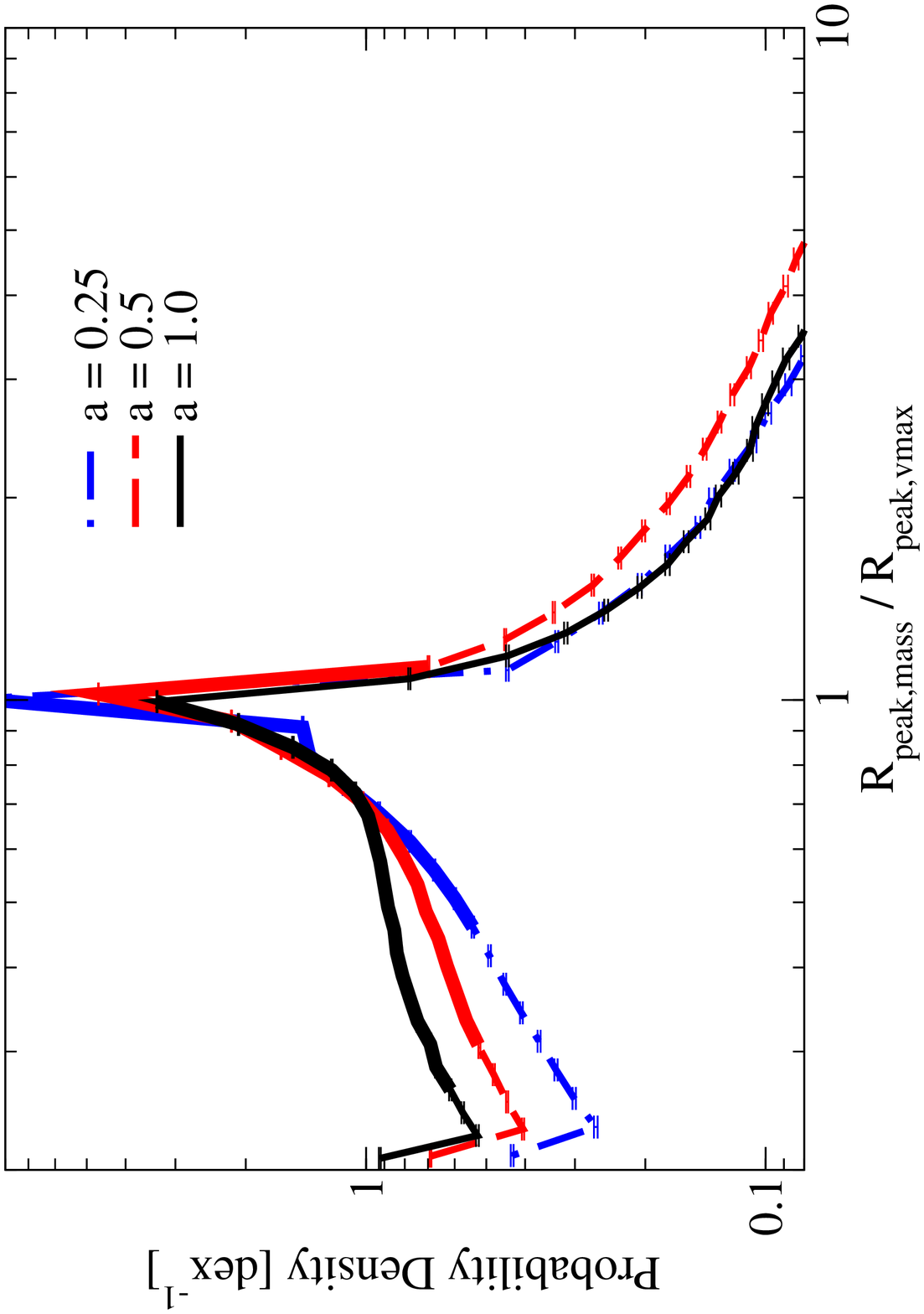}
\caption{\textbf{Top panel:} the evolution of $R_\mathrm{peak,mass} / R_\mathrm{acc,host}$ with redshift for the Bolshoi simulation with the \textsc{Rockstar} halo finder (analogous to Fig.\ \ref{f:rpeak2}).  \textbf{Double-width lines} correspond to the middle 68\% of the probability distributions.  \textbf{Middle panel}: conditional density plot of $R_\mathrm{peak,mass} / R_\mathrm{acc,host}$ as a function of satellite halo peak mass.  As with Fig.\ \ref{f:rpeak}, more massive satellites are more likely to be major mergers, which can keep accreting even after they are within the host's virial radius.  A smoothing kernel with a FWHM of 0.12 dex has been applied; individual high-mass satellite halos therefore appear as round blobs.  \textbf{Bottom panel}: histograms of  $R_\mathrm{peak,mass} / R_\mathrm{peak,vmax}$ as a function of host mass from $z=0$ to $z=3$.  Both radii are calculated as the distance to the satellite's host's most-massive progenitor at the time of peak mass and $\vmax$, respectively.  The central peak ($R_\mathrm{peak,mass}$ within 10\% of $R_\mathrm{peak,vmax}$) corresponds to $\sim 20\%$ of all halos; about 80\% of halos have $R_\mathrm{peak,mass} < R_\mathrm{peak,vmax}$, regardless of redshift.}
\label{f:mpeak}
\end{figure}

\subsection{The Radius of Peak Infall Mass}
\label{s:rmass}

The radius of peak infall mass  is simpler to interpret than the radius of peak $\vmax$.  We show the distribution and redshift evolution of $R_\mathrm{peak,mass}$ in the top panel of Fig.\ \ref{f:mpeak}.  At $z=0$, the median $R_\mathrm{peak,mass}$ is $1.8 R_\mathrm{acc,host}$, with a 68$^\mathrm{th}$-percentile range of $0.8$ to $4.1 R_\mathrm{acc,host}$.  This is broadly consistent with where the infalling halo's Hill radius shrinks to its virial radius, i.e., at $3^{1/3} \sim 1.4$ times $R_\mathrm{acc,host}$ (see, e.g., \citealt{Hahn09}).  The radius of peak infall mass becomes smaller relative to $R_\mathrm{acc,host}$ at higher redshifts partially due to the increased likelihood of surviving satellites being major mergers (\S \ref{s:rpeak_vmax}) and partially due to the effect of cosmological expansion on the gravitational force law, which makes it more difficult for dark matter to escape from the infalling halo at higher redshifts \citep{BehrooziUnbound}.

We show the conditional density plot of $R_\mathrm{peak,mass}/R_\mathrm{acc,host}$ at $z=0$ as well as the probability distribution of $R_\mathrm{peak,mass}/R_\mathrm{peak,vmax}$ in the bottom panels of Fig.\ \ref{f:mpeak}.  While a fraction of halos ($\sim$20\%) reach peak mass close to peak $\vmax$, the vast majority (80\%) have $R_\mathrm{peak,mass}$ less than $R_\mathrm{peak,vmax}$.  That said, we find that the halo mass increases by only 20\% on average after the time of peak $\vmax$ (for satellites at $z=0$).  This provides additional evidence that significant mergers after peak $\vmax$ are uncommon; otherwise, the mass would increase by a larger amount.  The peak mass is therefore most often set by smooth accretion and minor or very minor mergers (see also Fig.\ \ref{f:example}).

\begin{figure}
\plotgrace{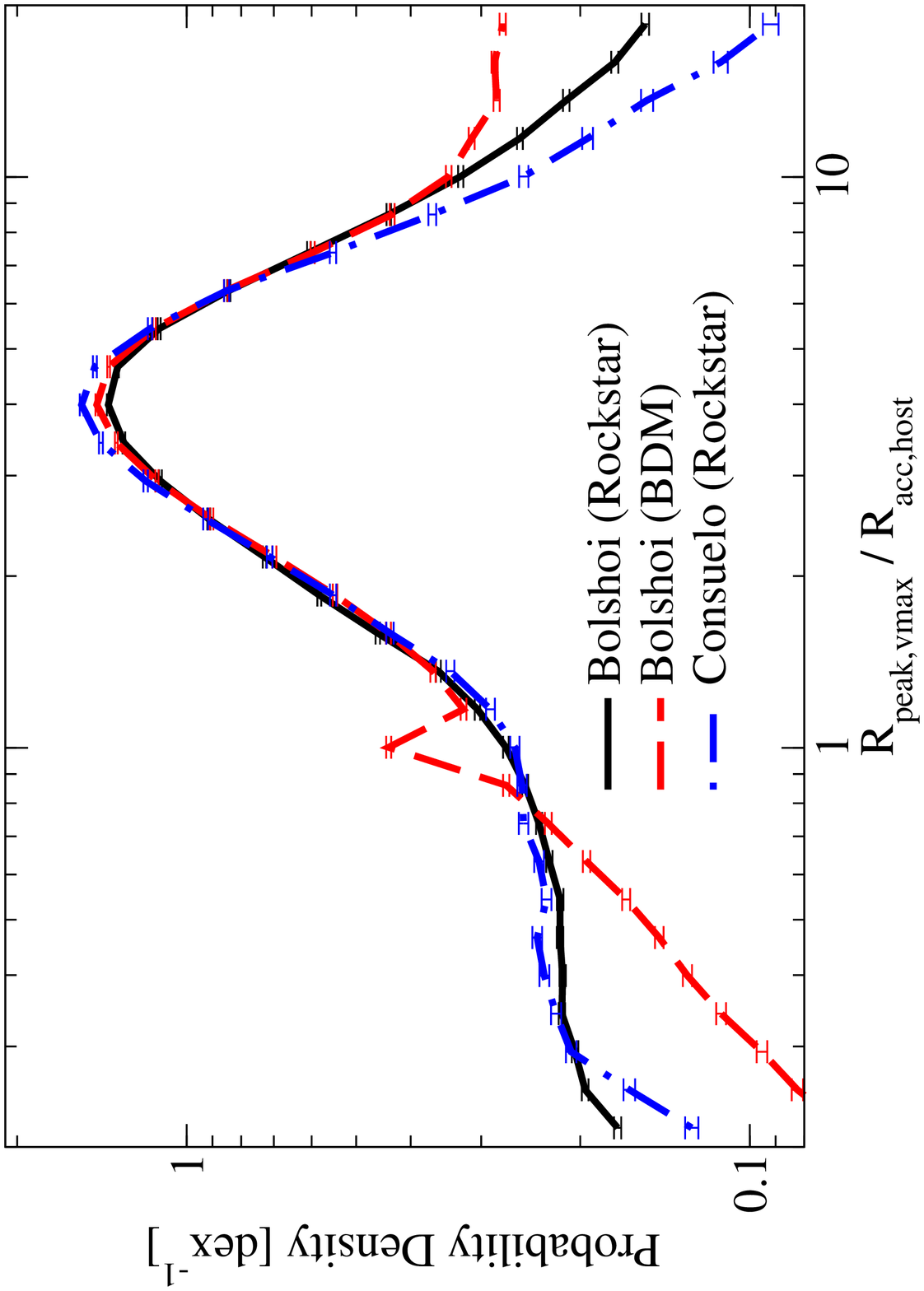}\\[-5ex]
\plotgrace{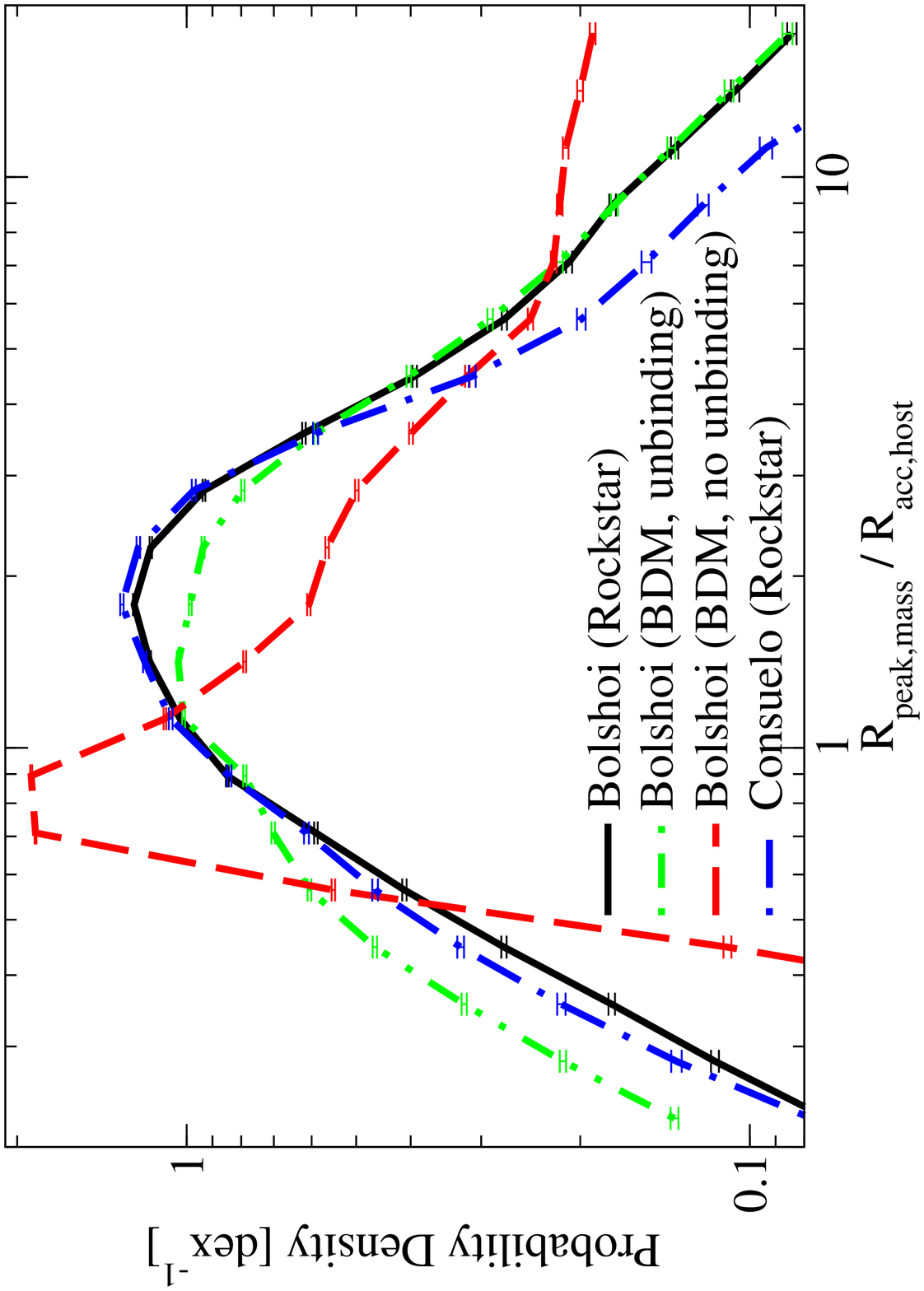}
\caption{\textbf{Top panel:} comparisons of the $R_\mathrm{peak,vmax} / R_\mathrm{acc,host}$ distribution for different combinations of simulations and halo finders at $z=0$.  There is universal agreement that the median $R_\mathrm{peak,vmax}$ is near $4$ $R_\mathrm{acc, host}$.  The differences within $R_\mathrm{acc,host}$ between \textsc{BDM} and \textsc{Rockstar} are due to the different techniques used for determining satellite particle membership; \textsc{BDM} applies a more conservative selection near the center of the host.  \textbf{Bottom panel:} comparison of $R_\mathrm{peak,mass} / R_\mathrm{acc,host}$ for different combinations of simulations and halo finders at $z=0$.  The main differences arise due to the gravitational unbinding choices for the different halo finders.  The current version of \textsc{BDM} does not do gravitational unbinding for host halos, meaning that infalling halos continue growing up to the virial radius of their eventual host due to increased overlap with its matter distribution.  On the other hand, \textsc{Rockstar} performs unbinding for all halos, more effectively separating the mass profile of the infalling halo from the mass profile of the host.  An older version of \textsc{BDM} which performed unbinding for all halos gives results which are more similar to \textsc{Rockstar}; however, this older version occasionally mis-assigned host particles to satellites, resulting in excess peak masses at low radii.  For this reason, we only discuss the \textsc{Rockstar} results for $R_\mathrm{peak,mass}$ in this paper.}
\label{f:mpeak2}
\end{figure}

\subsection{First Approach vs.\ Backsplash (Flyby) Halos}
\label{s:backsplash}

The plots in the previous sections do not specify whether infalling halos reach peak $\vmax$ or mass on their first approach or whether they pass through the virial radius of their eventual host and reach peak $\vmax$/mass afterwards.  These ``backsplash'' halos (also sometimes called ``flyby'' halos) would have very different mass accretion histories than halos on their first approach, which could have an important effect on their galaxy properties.  The typical backsplash radius ($2.5 R_\mathrm{vir,host}$ at $z=0$; \citealt{Gill05,Sinha12,Oman13,Wetzel13}) is smaller than the typical radius of peak $\vmax$, but larger than the typical radius of peak mass.  We find that the fraction of halos which reach peak $\vmax$ after backsplash is always less than 10\% (regardless of satellite mass, host mass, and redshift); for peak mass, this fraction is always less than 15\%.  These fractions are small, presumably because mass loss within the virial radius is very severe \citep{Tormen98b,Kravtsov04b,Knebe06}.

\subsection{Different Halo Finders and Simulations}

\label{s:hf_sims}

As shown in the top panel of Fig.\ \ref{f:mpeak2}, there is excellent agreement between all halo finders and simulations for the probability distribution of $R_\mathrm{peak,vmax}$, except in the low-probability tails of the distribution.  The largest minor difference comes between the \textsc{Rockstar} and \textsc{BDM} halo finders; because \textsc{BDM} turns on unbinding for satellite halos only, a disproportionate number of infalling halos peak in $\vmax$ just outside the virial radius of the larger host halo.

Halo mass is more ambiguous to determine than circular velocity, especially for satellite halos \citep{Knebe11,Onions12,Knebe13}.  While the different simulations agree very well on the probability distribution for $R_\mathrm{peak,mass}$, the bottom panel of Fig.\ \ref{f:mpeak2} suggests that different halo finders can give different results.  As mentioned in \S \ref{s:hf}, the \textsc{Rockstar} halo finder performs unbinding for all halos, whereas \textsc{BDM} performs unbinding only for satellite halos.  Because the mass distribution of the eventual host halo extends well beyond its virial radius \citep[e.g.,][]{Busha03}, particles within the virial radius of the infalling halo can include significant contamination from the host mass distribution \citep{BehrooziUnbound}.  This difference means that infalling halos in \textsc{BDM} will ``accrete'' unbound matter all the way into the virial radius of the eventual host.  As our primary concern in this study is the effect on galaxy formation, and because the much hotter host halo gas is unlikely to be accreted onto the galaxy in the infalling halo, we have only presented results from the \textsc{Rockstar} halo finder in the remainder of this section.  Nonetheless, when using an older version of \textsc{BDM} which performs unbinding on all halos, we find results which are more similar to \textsc{Rockstar} for the probability distribution of $R_\mathrm{peak,mass}$ (Fig.\ \ref{f:mpeak2}, bottom panel).

\begin{figure}
\plotgrace{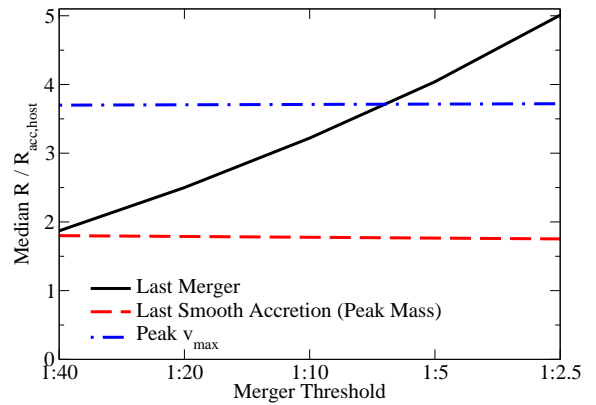}
\caption{The median clustercentric distance ($R / R_\mathrm{acc,host}$) for infalling halos' last mergers as a function of the minimum merger ratio threshold, compared to median clustercentric distances for peak $\vmax$ and mass.  The last merger larger than 1:40 typically happens before peak mass, i.e., the end of smooth accretion.  However, peak $\vmax$ is often set by the last $\sim$1:5 merger, and smaller mergers continue after the infalling halo reaches peak $\vmax$.}
\label{f:mean_r}
\end{figure}

\section{Discussion}

\label{s:discussion}

In this section, we discuss how our results on clustercentric distances (\S\ref{s:rpeak}; summarized in Fig.\ \ref{f:mean_r}) affect the theoretical interpretation of galaxy quenched fractions in \S \ref{s:observed}, galaxy morphologies in clusters in \S \ref{s:morphologies}, and abundance matching on peak $\vmax$ in \S \ref{s:abundance}.

In translating the results on halos  to effects on visible galaxies, some care is required.  Most importantly, the increased binding energy of halo inner regions when baryons are included may mean that satellite galaxies will not have visible changes (e.g., in their color or star formation rate) even after much of their surrounding dark matter halo is gone.  Secondly, halo mass merger ratios are not the same as galaxy mass merger ratios.  Due to the shape of the stellar mass---halo mass relation (e.g., \citealt{BWC12}), merger ratios are compressed for galaxies above $10^{10.5}\Msun$ in stellar mass---i.e., minor mergers in halo mass will be major mergers in galaxy mass.  On the other hand, merger ratios are expanded for galaxies below $10^{10.5}\Msun$ in stellar mass, so that even major mergers in halo mass can become minor mergers in galaxy mass.  Finally, halos may lose gas due to ram pressure stripping before they lose dark matter from tidal stripping (e.g., \citealt{Bahe13}).  Gas profiles in simulations remain sensitive to the exact feedback mechanisms employed \citep{Scannapieco12}, somewhat complicating the interpretation of these results.  However, in the discussion that follows, we treat the tidal stripping radius as a lower limit on the gas stripping radius.

\subsection{Interpretation of Quenched Fraction Profiles near Clusters}

\label{s:observed}

As discussed in \S \ref{s:intro}, it is an open question whether satellite galaxies are quenched in merger-triggered events or through gas stripping processes.  Recently, \cite{Wetzel13} found enhancements in the quenched fraction of central galaxies out to 2.5 $\rvir$ around clusters at $z\sim0$, with a small effect seen out to $5\rvir$.  This excess could be accurately fitted if infalling galaxies become quenched starting at the virial radius of clusters \citep{Wetzel13}; increased quenching beyond the virial radius would correspond to ``flyby'' galaxies (which enter and then leave the cluster virial radius) from the main cluster as well as from correlated structure.

If it is true that galaxies are quenched only after entering the virial radii of larger hosts, the fact that $>$1:5 mergers end beyond 4$R_\mathrm{vir,host}$ for most halos suggests that merger-induced feedback (such as merger-triggered AGN feedback) is less likely to be the cause of quenching.  As noted earlier, galaxy mass merger ratios will not be the same as halo mass merger ratios; we use the stellar mass--halo mass relation in \cite{BWC12} to convert between the two.  For a $10^{11}\Msun$ (stellar mass) host galaxy, a 1:10 merger in galaxy mass would correspond to a 1:40 merger in halo mass.  As these mergers end at about the same clustercentric distance as where tidal mass stripping begins (Fig.\ \ref{f:mpeak2}), it is plausible that merger-triggered AGN activity could heat any existing cold star-forming gas, and that tidal forces would prevent any new cold gas from being accreted.  Thus, we cannot exclude this mechanism from increasing the fraction of quenched $10^{11}\Msun$ galaxies (which is already large even for field galaxies; \citealt{Salim07}).   However, for a $10^{10}\Msun$ host galaxy, a 1:10 merger in galaxy mass would correspond to a 1:3 merger in halo mass; these mergers would be expected to end at $\sim 5 R_\mathrm{vir,host}$.  The large spatial offset between the end of mergers and the onset of quenching suggests that mergers are less likely to cause immediate quenching in lower-mass galaxies.

Of course, merger-triggered AGN activity \textit{with a significant time delay} between a galaxy merger and the onset of feedback may also be able to explain the \cite{Wetzel13} radial quenched fractions.  A simple test for this model would be to exclude all potential flyby galaxies (i.e., galaxies within $2.5\rvir$ of any cluster) and to recalculate radial quenched fractions near clusters.  Most satellite galaxies today had their last major merger well beyond $4R_\mathrm{vir,host}$ (Fig.\ \ref{f:mean_r}), so unless halos at that distance know in advance whether they will fall into clusters or not,  merger-triggered quenching would imply that some quenching enhancement should appear at $>4R_\mathrm{vir,host}$ regardless of the time delay.

As an example, \cite{Geha12} report that the vast majority of quenched dwarf galaxies (stellar masses between $10^7\Msun$ and $10^9\Msun$) are within 2 virial radii of the nearest massive host (stellar mass greater than $2.5\times10^{10}\Msun$); the quenched fraction beyond $4\rvir$ ($\sim 0.2\%$) is only negligibly enhanced over the field value ($<0.06\%$).  This can be taken as additional evidence that merger-induced quenching is likely not dominant in dwarf galaxies.  While this result was far from unexpected, tests on quenched fractions for $10^{10.5}\Msun$ and larger galaxies may yield more interesting conclusions (Behroozi et al., in prep.).

\subsection{Galaxy Morphologies in Clusters}
\label{s:morphologies}

Satellite galaxies are more likely to be ellipticals or spheroidals than are field galaxies (see, e.g., \citealt{Blanton09} for a review).  One reason for the comparative lack of spirals in clusters is that dry mergers may convert spirals into ellipticals; it is also possible that pseudobulge formation or passive fading of disk stars could convert some spirals into spheroidal/S0 galaxies \citep[and references therein]{Kormendy04,Weinmann09,Kormendy12}.  Additionally, spiral disks may also be heated and dispersed more easily in clusters \citep{Kormendy12}.

As noted in the previous section, a given merger ratio in galaxy mass requires a more massive merger ratio in halo mass for galaxies less than $\sim 10^{10.5}\Msun$ in stellar mass.  For example, for $10^{10.5}\Msun$ stellar mass galaxies, 1:40 mergers in galaxy mass correspond to 1:10 mergers in halo mass \citep{BWC12}, which end for most infalling galaxies at $3.2\rvir$ (Fig.\ \ref{f:mean_r})---i.e., prior to the end of mass accretion and star formation (as inferred by the quenched fraction of galaxies near clusters).  Moreover, because most of the corresponding merging galaxies will have stellar masses in the range $10^{9}\Msun$ to $10^{10}\Msun$, they will tend to be gas-rich \textit{wet} mergers, rather than dry ones \citep{Erb06,Baldry08,Stewart09,Wei10}. As wet mergers will lead to disks reforming, it is difficult to explain morphology changes in $<10^{10.5}\Msun$ satellite galaxies due to mergers alone \citep[see also][]{Stewart09}.

For infalling galaxies larger than $\sim 10^{10.5}\Msun$, the situation is reversed.   For a $10^{11.0}\Msun$ stellar mass spiral, a 1:10 merger in terms of halo mass will on average be a 1:2.5 merger in galaxy mass; a 1:40 merger in halo mass will still be a 1:10 merger in galaxy mass.  These latter mergers will happen much closer to the cluster virial radius (Fig.\ \ref{f:mean_r}).  These mergers will also tend to be \textit{dry} mergers, due both to the lack of mass accretion close to the cluster as well as the lower base gas fractions in the merging galaxies \citep{Erb06,Baldry08,Stewart09,Wei10}.  This would mean that dry mergers may contribute to reducing the remaining fraction of spirals at these larger galaxy masses.

\subsection{Implications for Abundance Matching}

\label{s:abundance}

Recently, \cite{Reddick12} found that $\vpeak$ is a better proxy for stellar mass than most other halo properties (including $\vmax$ at accretion, peak mass, and mass at accretion) for modeling autocorrelation and conditional luminosity functions.  However, interpreting \textit{why} $\vpeak$ is the best proxy of those considered is more difficult.  As shown in \S \ref{s:last_mm} and Appendix \ref{a:timing}, $\vpeak$ is often set by a merger, which results in only a transient increase in $\vmax$.  Indeed, galaxies continue to accrete gas (\S \ref{s:rmass}) and form stars well within the distance where $\vpeak$ is set \citep{Wetzel13}.

The primary constraint from observations of the correlation function and the conditional luminosity function shown in \cite{Reddick12} is on the fraction of galaxies that are satellites, as a function of galaxy abundance (ranked by luminosity or stellar mass).  It is possible that the use of $\vpeak$ as compared to other proxies (e.g., $\vmax$ at the time of accretion or peak mass) partially compensates for premature loss of satellite halos in simulations \citep[see also][]{MillOrphan}.  At fixed peak halo mass, satellites have higher $\vpeak$ than centrals do \citep{Reddick12}, which boosts the satellite clustering signal.  However, keeping satellite halos around longer will also boost the satellite clustering signal in a largely degenerate manner \citep{MillOrphan,wetzel-09,Reddick12}.  This degeneracy may be broken by noting that abundance matching on $\vpeak$ gives different implications for mean galaxy growth near clusters as compared to, e.g., abundance matching on peak mass (see also Appendix \ref{a:timing}).  Thus, testing abundance matching models against galaxy number counts in \textit{annuli} around clusters (e.g., $R_\mathrm{vir,host} < R < 2R_\mathrm{vir,host}$) where premature satellite halo loss in simulations is less significant may better constrain which halo properties best correspond to galaxy properties.

We note, however, that no proxy for stellar mass which is fixed at a single epoch (e.g., peak mass or $\vpeak$) will capture the orbit-dependent effects of mass stripping (both gas and dark matter) from satellite galaxies.  These effects will strongly influence the star formation rates of individual satellite galaxies.  Instead, it may be worthwhile for future abundance matching studies to model satellites' stellar mass growth according to their unique mass accretion / stripping histories \citep{Behroozi13}. 

\section{Conclusions}
\label{s:conclusions}

We have investigated the radii at which host halos begin to influence their nearby environment, quantifying the range of distances from the host center that infalling halos experience their last mergers, peak mass, and peak circular velocity.
The main results may be summarized as follows:

\begin{enumerate}
\item Peak $\vmax$ for halos is often set by their last 1:5 or larger merger (\S \ref{s:last_mm}, Appendix \ref{a:timing}); the merger causes a temporary $\sim$14\% boost in $\vmax$ which lasts for $\sim2$ dynamical times (Appendix \ref{a:timing}).

\item Peak $\vmax$ for infalling halos occurs at a median distance of $3.7$ times the virial radius ($\rvir$) of the eventual host halo at $z=0$ (\S \ref{s:rpeak_vmax}); at higher redshifts, it occurs closer relative to the host radius ($\sim3\rvir$ at $z=1$ and $\sim 2 \rvir$ at $z=3$).
\item Peak mass for infalling halos occurs at $\sim 1.8\rvir$ of the final host halo at $z=0$, which is near the radius at which tidal forces from the host make orbits at the virial radius of the infalling halo unstable ($\sim 1.4\rvir$; \S \ref{s:rmass}).  Peak mass corresponds to the end of smooth accretion and very minor mergers (\S \ref{s:rmass}).
\item Halo finders which do not perform gravitational unbinding on all halos will find that infalling halos ``grow'' until they become satellites (\S \ref{s:rmass}).  This is due to increasing contamination from particles associated with the eventual host halo; the associated gas would likely be too hot to condense onto the infalling halo's central galaxy.
\item It is very unlikely for halos to reach peak mass or $\vmax$ after passing within the virial radius of a larger halo (\S \ref{s:backsplash}).
\item Based on the radial profile of quenched fractions near clusters in \cite{Wetzel13}, it is plausible that galaxy quenching near clusters is correlated with mass stripping.  Merger-induced quenching would likely result in more extended quenching profiles than seen in \cite{Wetzel13} (\S \ref{s:observed}).
\item For low stellar mass ($<10^{10.5}\Msun$) satellite galaxies in clusters, it is likely that morphological transitions from spiral to spheroidal or elliptical shapes are not due to dry mergers (\S \ref{s:morphologies}).  However, dry mergers may explain some morphological evolution in larger (stellar mass $>10^{10.5}\Msun$) galaxies.
\item The success of $\vpeak$ as a proxy for stellar mass in abundance matching studies \citep[e.g.,][]{Reddick12} may be due primarily to it better reproducing the observed satellite fraction in the simualtions studied, rather than to a deeper physical connection with star formation (\S \ref{s:abundance}).
\end{enumerate}

\acknowledgements
PB and RHW received support from an HST Theory grant; program number HSTAR- 12159.01-A was provided by NASA through a grant from the Space Telescope Science Institute, which is operated by the Association of Universities for Research in Astronomy, Incorporated, under NASA contract NAS5-26555.   PB was also partially supported by
a Giacconi Fellowship through the Space Telescope Science Institute. OH acknowledges support from the Swiss National Science Foundation
(SNF) through the Ambizione fellowship. We also appreciate many helpful discussions and comments from Charlie Conroy, Andrew Hearin, Avi Loeb, Houjun Mo, and Ramin Skibba.

%\newpage

\bibliography{master_bib}

\appendix

\begin{figure}
\plotminigrace{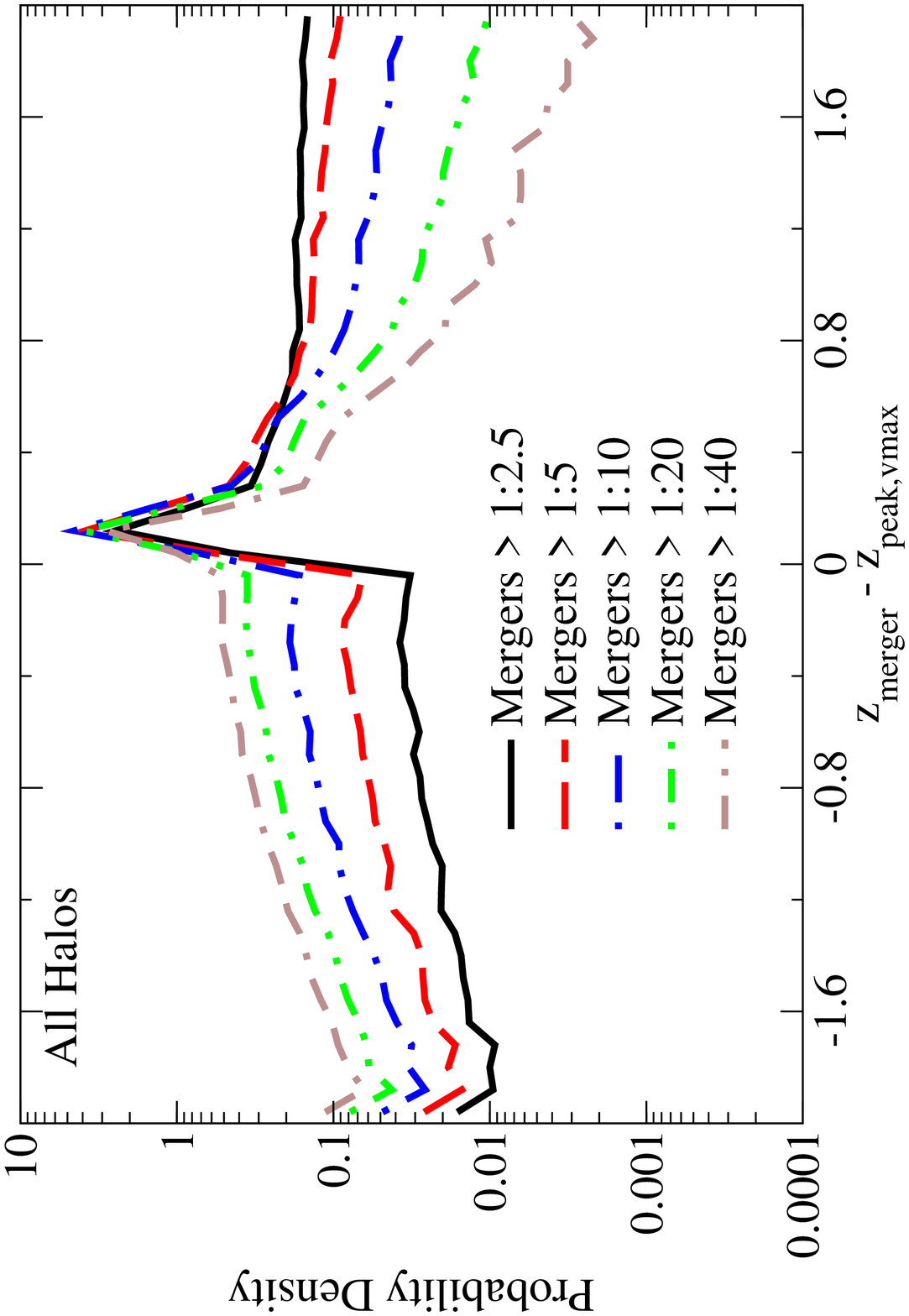}\plotminigrace{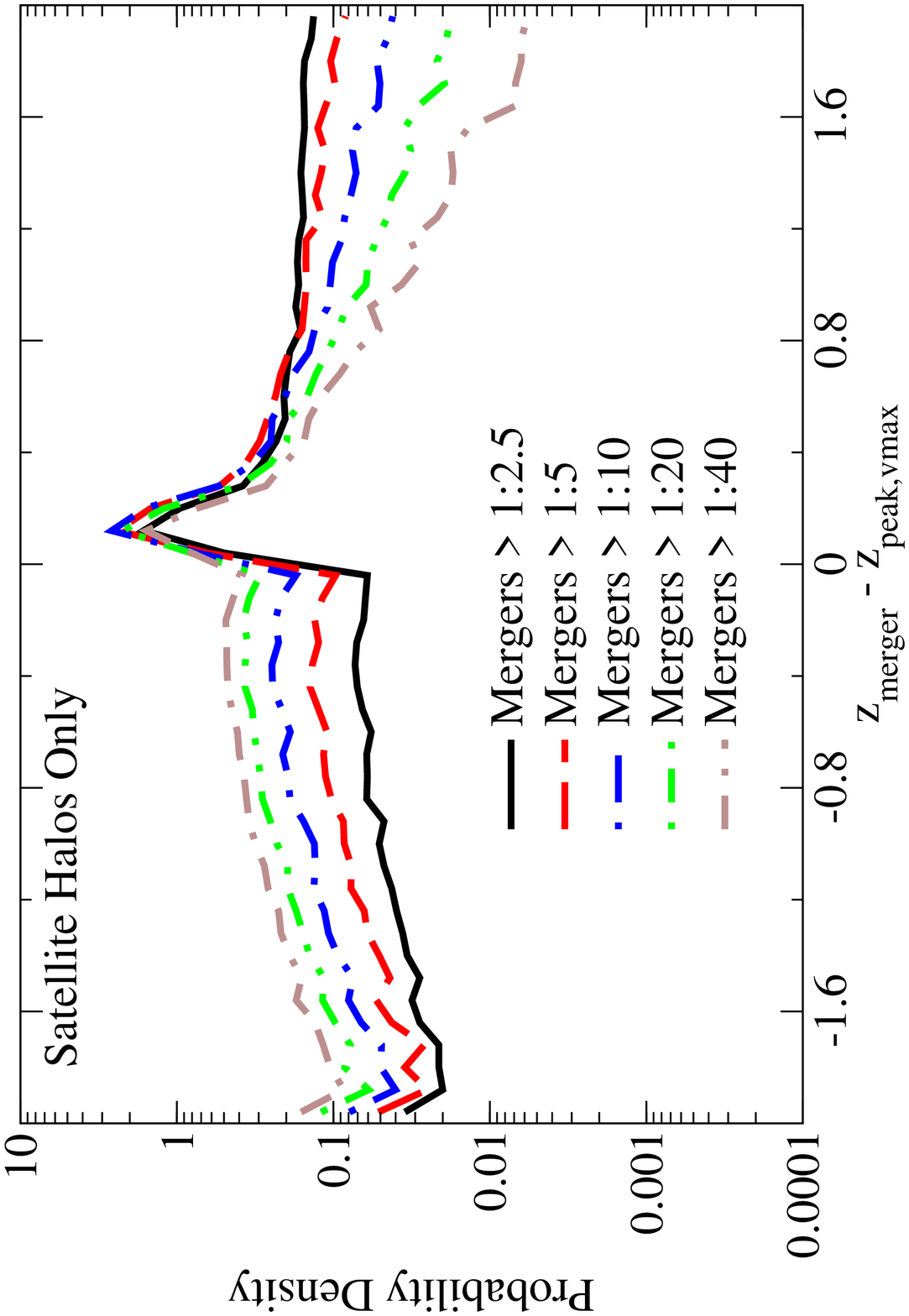}\\[-5ex]
\plotminigrace{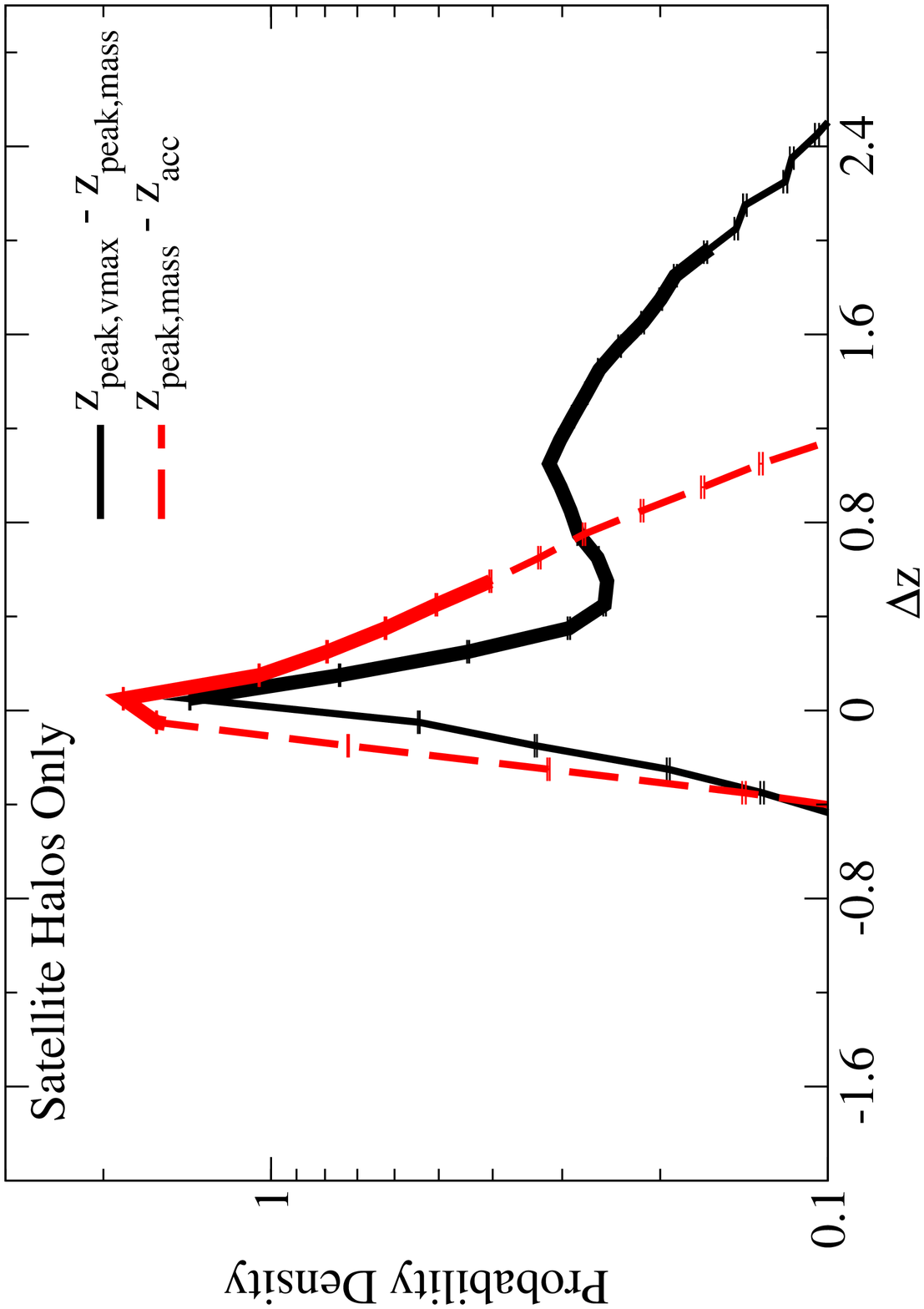}\plotminigrace{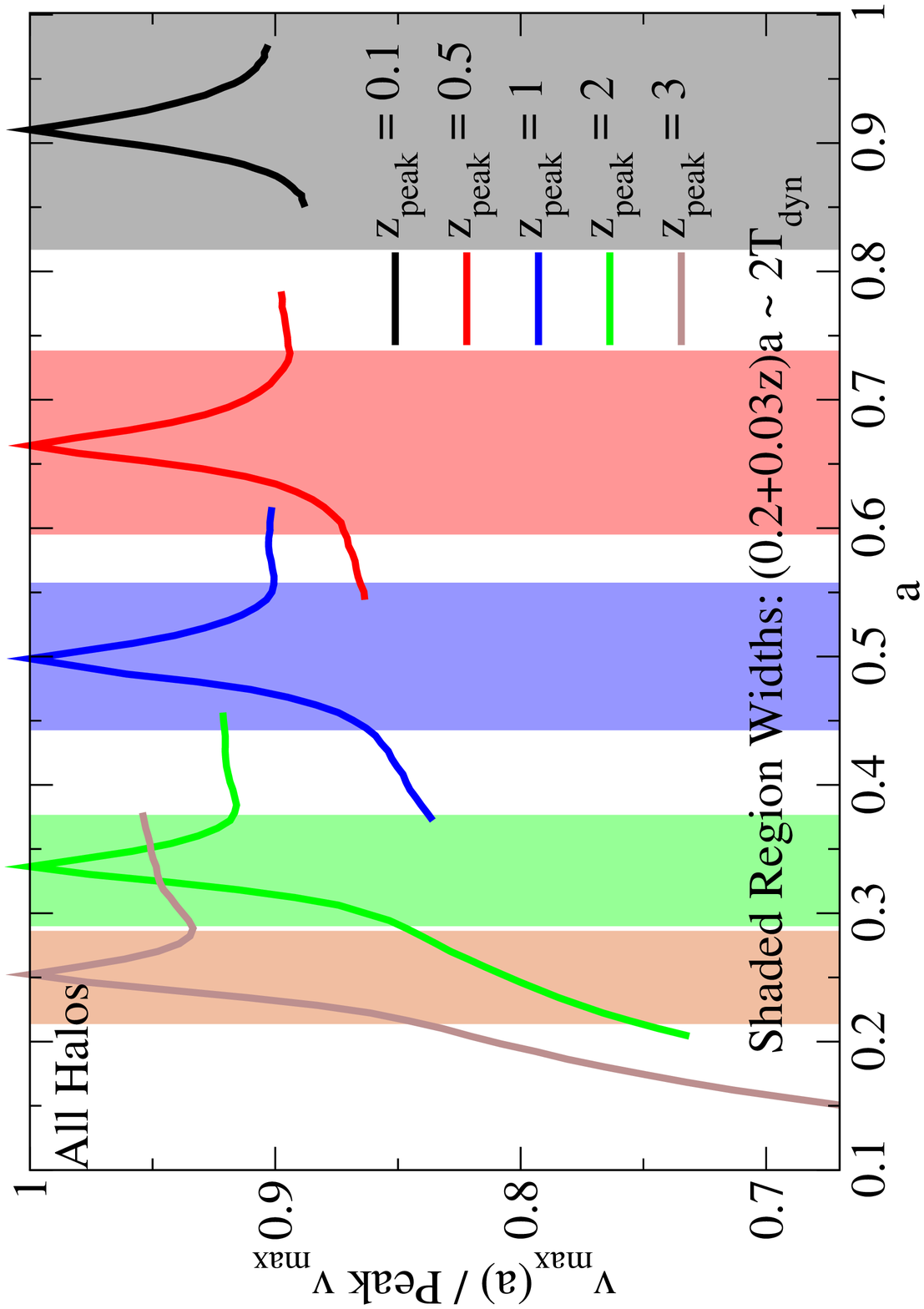}
\vspace{-5ex}
\caption{\textbf{Top-left} panel: redshift of last merger compared to redshift of peak $\vmax$ for all halos, as a function of merger threshold, for all halos.  A very similar effect is seen in Fig.\ \ref{f:merger_hist}, where mergers are highly correlated with peak $\vmax$.  \textbf{Top-right} panel: same, but counting satellite halos only.  If anything, satellite halos show a slightly weaker correlation than all halos taken together.  \textbf{Bottom-left} panel: comparisons between redshifts of peak $\vmax$ ($z_\mathrm{peak,vmax}$), peak mass ($z_\mathrm{peak,mass}$), and first accretion ($z_\mathrm{acc}$) for satellite halos.  Accretion follows soon after peak mass; however, a significant fraction of infalling halos have large delays between peak $\vmax$ and peak mass.  \textbf{Double-width lines} show the middle 68\% of the probability distributions.  \textbf{Bottom-right} panel: median ratios of $\vmax$ to peak $\vmax$ for all halos which have a peak in $\vmax$ occurring at $z=0.1,0.5,1,2$ or $3$ (see text).  These peaks in halo $\vmax$ histories last for roughly two dynamical times.  All panels show results from the Bolshoi simulation with the \textsc{Rockstar} halo finder.}
\label{f:timing}
\end{figure}

\section{Timing of Mergers, Peak Mass, and Peak $\vmax$}

\label{a:timing}

As noted in \S\ref{s:last_mm}, the radius of the last $>$1:5 merger is highly correlated with the radius of peak $\vmax$.  It is interesting to test whether host halos also experience a similar effect.  Because the clustercentric distance is less meaningful for host halos, we compare the redshift of the last merger to the redshift of peak $\vmax$ for host and satellite halos at $z=0$ in the top-left panel of Fig.\ \ref{f:timing}.  Compared to the probability distribution for satellites only (Fig.\ \ref{f:timing}, top-right panel), the correlation between mergers and peak $\vmax$ seems slightly stronger for all halos than it does for satellite halos alone.

It is also interesting to consider the timing for when infalling halos reach peak $\vmax$, peak mass, and the virial radius of their eventual host halos, as shown in the bottom-left panel of Fig.\ \ref{f:timing} for satellite halos at $z=0$.  Infalling halos generally become accreted shortly after reaching peak mass ($\Delta z < 0.5$), but it takes them much longer after peak $\vmax$ to reach peak mass (68$^\mathrm{th}$-percentile range: $0<\Delta z<2$).  As noted in Fig.\ \ref{f:mean_r}, infalling halos reach peak $\vmax$ at much larger clustercentric distances than they reach peak mass.  Because infalling halos accelerate as they approach their eventual host halos, infalling halos will spend much more time at large radii than they will at smaller ones.  This implies that halo peak $\vmax$ will remain fixed for much longer than peak mass, which will result in different stellar populations for galaxies near clusters as inferred from abundance matching on peak $\vmax$ compared to abundance matching on peak mass (\S \ref{s:abundance}).

We also consider how long transient peaks in $\vmax$ last.  We first select all halos from the Bolshoi simulation which have a transient peak in their $\vmax$ history, defined as reaching a $\vmax$ which is not exceeded for the next 5 timesteps.  Then, for these halos, we record the ratio of $\vmax$ to the transient peak $\vmax$ for 20 timesteps before and after the peak.  To ensure that the $\vmax$ history is well resolved even at high redshifts, we exclude halos with masses below $5\times 10^{10}\Msun$.  The bottom-right panel of Fig.\ \ref{f:timing} shows the median ratio of $\vmax$ to the transient peak $\vmax$ for halos as a function of the transient peak redshift.

From these ratios, we find that the peak halo $\vmax$ during a transient is typically 14\% higher than the average of the halo's $\vmax$ before and after the transient, regardless of redshift.  In addition, we find that the following fitting formula matches the typical width of transient peaks in halo $\vmax$ histories:
\begin{equation}
\label{e:transient}
\Delta a = (0.2 + 0.03 z_\mathrm{peak}) a_\mathrm{peak}
\end{equation}
Where $\Delta a$ is the difference in scale factor between the beginning and end of the transient, $z_\mathrm{peak}$ is the redshift of the transient peak, and $a_\mathrm{peak}$ is $(1+z_\mathrm{peak})^{-1}$.  For $z\sim0.5$ to $z\sim3$, $\Delta a$ is approximately two dynamical times; however, higher-redshift transients typically last for slightly more dynamical times than lower-redshift ones.  Smoothing $\vmax$ histories over several dynamical times will reduce the significance of these peaks; future work will test if these smoothed histories yield more physically realistic results for, e.g., abundance matching techniques.

\end{document}